\documentclass[twocolumn, numberedappendix]{aastex62}

\usepackage{caption}
\usepackage{amstext}
\usepackage{amsmath}
\usepackage{amssymb}
\usepackage{graphicx}
\usepackage{upgreek}

\newcommand{\beq}{\begin{equation}}
\newcommand{\eeq}{\end{equation}}

\graphicspath{{./}{figures/}}

\begin{document}

\title{The Evolution of Binaries under the Influence of Radiation-Driven Winds from a Stellar Companion}

\correspondingauthor{Sophie Lund Schr\o der}
\email{sophie.schroder@nbi.ku.dk}

\author[0000-0003-1735-8263]{Sophie Lund Schr\o der}
\affiliation{Niels Bohr Institute, University of Copenhagen, Blegdamsvej 17, 2100 Copenhagen, Denmark}
\affiliation{Center for Astrophysics | Harvard \& Smithsonian, 60 Garden Street, Cambridge, MA, 02138, USA}

\author[0000-0002-1417-8024]{Morgan MacLeod}
\affiliation{Center for Astrophysics | Harvard \& Smithsonian, 60 Garden Street, Cambridge, MA, 02138, USA}

\author[0000-0003-2558-3102]{Enrico Ramirez-Ruiz}
\affiliation{Department of Astronomy and Astrophysics, University of California, Santa Cruz, CA 95064, USA}
\affiliation{Niels Bohr Institute, University of Copenhagen, Blegdamsvej 17, 2100 Copenhagen, Denmark}

\author[0000-0002-6134-8946]{Ilya Mandel}
\affiliation{Monash Centre for Astrophysics, School of Physics and Astronomy, Monash University, Clayton, Victoria 3800, Australia}
\affiliation{The ARC Center of Excellence for Gravitational Wave Discovery -- OzGrav, Australia}
\affiliation{Birmingham Institute for Gravitational Wave Astronomy and School of Physics and Astronomy, University of Birmingham,Birmingham, B15 2TT, United Kingdom}

\author[0000-0003-1474-1523]{Tassos Fragos}
\affiliation{Geneva Observatory, University of Geneva, Chemin Pegasi 51, 1290 Versoix, Switzerland}

\author[0000-0003-4330-287X]{Abraham Loeb}
\affiliation{Center for Astrophysics | Harvard \& Smithsonian, 60 Garden Street, Cambridge, MA, 02138, USA}

\author[0000-0001-5256-3620]{Rosa Wallace Everson}
\altaffiliation{NSF Graduate Research Fellow}
\affiliation{Department of Astronomy and Astrophysics, University of California, Santa Cruz, CA 95064, USA}

\begin{abstract}
Interacting binaries are of general interest as laboratories for investigating  the physics of accretion, which gives rise to the bulk of high-energy radiation in the Galaxy.  They allow us to probe stellar evolution processes that cannot be studied in single stars. Understanding the orbital evolution of binaries is essential in order to model the formation of compact binaries. Here we focus our attention on studying  orbital evolution driven by angular momentum loss through stellar winds in massive binaries. We run a suite of hydrodynamical simulations of binary stars hosting one mass losing star with varying wind velocity, mass ratio, wind velocity profile and adiabatic index, and compare our results to analytic estimates for drag and angular momentum loss. We find that, at leading order, orbital evolution is determined by the wind velocity and the binary mass ratio. Small ratios of wind to orbital velocities and large accreting companion masses result in high angular momentum loss and a shrinking of the orbit. For wider binaries and binaries hosting lighter mass-capturing companions, the wind mass-loss becomes more symmetric, which results in a widening of the orbit. We present a simple analytic formula that can accurately account for  angular momentum losses and changes in the orbit, which depends on the wind velocity and mass ratio. As an example of our formalism,  we compare the effects of tides and winds in driving the orbital evolution of high mass X-ray binaries, focusing on Vela X-1 and Cygnus X-1 as examples.
\end{abstract}

\section{Introduction}

Low-mass main sequence stars like our Sun lose only a tiny fraction of their mass through stellar winds over their main sequence evolution. In contrast, massive stars can produce stellar winds a billion times stronger and, during their much swifter evolution, will shed up to half of their mass. This large  mass loss can have profound consequences for the evolution of the star alone, but even more so when the star is a member of a binary system \citep{2015HiA....16...51V,2015ASSL..412..199W,2018A&A...615A.119V,2013ARA&A..51..269D}. Winds interacting with the binary induce drag forces strong enough to change the orbit of the stars \citep{1977MNRAS.179..265L,1993ApJ...410..719B,2018MNRAS.473..747C,2018A&A...618A..50S,2019A&A...626A..68S,2019A&A...629A.103S}. The strength of the drag forces can determine whether the stars go through a dynamically unstable mass transfer phase and either merge or form a compact binary, or grow so far apart that they evolve  essentially as single stars.

In a seminal work \citet{2005A&A...441..589J} attempted to build a general framework for the wind-driven drag force in binary systems. In order to take into account the complicated wind launching mechanism, \cite{2005A&A...441..589J} considered three models that approximate mechanically injected, thermally driven, and radiation-driven winds. The key result derived  by \citet{2005A&A...441..589J}, which has been confirmed in follow up studies  \citep[e.g.][]{2018A&A...618A..50S,2019A&A...626A..68S}, is that the resultant torque on the binary is most sensitively  dependent on the wind velocity.

Radiation-driven winds are expected for giant stars, where radiation is absorbed by dust grains, and for massive stars, where Compton scattering of electrons and line absorption are the primary wind acceleration mechanisms  \citep{2008A&ARv..16..209P}.
A challenging aspect of the simulating wind interactions in binary systems is modeling the launching and subsequent acceleration of the  wind. This is particularly true in the case of radiation-driven winds, where multi-frequency radiation-hydrodynamic simulations remain very computationally demanding for three-dimensional models. Realistic massive-star winds are likely clumpy \citep[e.g.][]{2020MNRAS.493..447C,2020A&A...643A...9E}, and may be inhibited by photoionization feedback from an accreting compact object companion \citep[e.g.][]{1990ApJ...356..591B,2016A&A...589A.102B,2018MNRAS.475.3240E,2018A&A...620A.150K}.

For the purposes of modeling wind effects on binary orbital evolution, the inclusion of all of these potential processes is unrealistically complex. Approximate, phenomenological models can be adopted in lieu of a full treatment. One simple model involves partially or completely turning off the gravitational influence  of the donor star as a function of radial distance, an idea that can be traced to the early work of \citet{1970ApJ...159..879L,1975ApJ...195..157C}.
This approximation results in a $\beta$-law profile, $v_w/v_\infty \approx (1-r_s/r)^\beta$, where $\beta = 0.5$ and $r_s$ is the sonic radius marking the transition from a subsonic to a supersonic wind. More recent studies have suggested values where $\beta \sim 0.7$ or non-monotonic profiles \citep{2008A&A...492..493M,2021A&A...647A.151P}. On the other hand the simplicity of implementing $\beta = 0.5$ winds via a reduced gravitational force in three dimensional studies makes it  more tractable when studying the wind interaction in the context of a binary system and hence we choose to adopt this simplification.

In this paper, we use hydrodynamic simulations to study the effects of stellar winds on a binary whose separation is compact, such that it lies within the region where the wind is still accelerating toward its terminal velocity. We examine winds with different terminal velocities, winds with different velocities measured at the binary separation but identical asymptotic velocities, and winds with matched velocities but differing acceleration profiles in order to assess the relative importance of these differing properties. Through these experiments, we attempt to ascertain what is the fundamental characteristic of the velocity of winds that determines their effect on their host binary.  

In Section \ref{sec:theory} we derive relations for changes in binary star orbits for varying wind velocities and binary mass ratios. 
In Section \ref{sec:hydro} we describe the setup in \texttt{Athena++}. In Section \ref{sec:Results} we present the results of our simulations 
with different wind velocity, mass ratio and wind acceleration prescriptions. 
In Section \ref{sec:discussion} we compare our results to previous work and
present a discussion of our salient findings. Finally, in Section  \ref{sec:conlusion}, we describe how we plan to use the simulations to build a model for binary evolution including wind gas drag and provide our conclusions.

\section{Theoretical background}
\label{sec:theory}

In this Section we describe how a binary orbit changes due to mass and angular momentum loss carried by stellar winds. We arrive at the dimensionless parameter $\gamma_{\rm drag}$, which describes the orbital angular momentum lost due to interaction between the gaseous wind and the stars. We then use the analytic theory of Bondi-Hoyle-Lyttleton (BHL) drag to derive an expected value of $\gamma_{\rm drag}$, which will serve as a baseline for comparison to our hydrodynamic simulations in Section \ref{sec:Results}.

\subsection{Orbital Angular Momentum}\label{sec:theory_orbit}

 In a circular-orbit binary of two non-spinning stars with masses $M_1$ and $M_2$, reduced mass $\mu$, total mass $M$, semimajor axis $a$,  the total angular momentum is given by
\begin{equation}
J = \mu  v_{\rm orb} a =  \sqrt{ \frac{G (M_1 M_2)^2 a}{M}   }
\label{eq:orb_L}
\end{equation}
\noindent where $v_{\rm orb}=  \sqrt{ GM / a}$ is the Keplerian orbital velocity. Changes in $M_1$, $M_2$ and $a$ thus can all change the angular momentum of  the orbit. To understand how the orbit evolves when each of these parameters is altered, it is customary to take the time derivative of the orbital angular momentum squared, 
\begin{equation}
\frac{\dot{\left(J^2 \right)}}{J^2} =  2 \frac{\dot{J}}{J} = 2 \left(\frac{\dot{M_1}}{M_1} +  \frac{\dot{M_2}}{M_2} \right) - \frac{\dot{M}}{M} + \frac{\dot{a}}{a}.
 \label{eq:Jdot}
\end{equation}
A parallel analysis can be carried out including terms for orbits of non-zero eccentricity or stellar component spin. 

\subsection{A Single Mass-losing Star}

We now simplify  equation \eqref{eq:Jdot} because we focus in this paper on how the binary separation evolves when star $M_1$ is losing mass through a stellar wind $\dot{M}_1 < 0$. Therefore, we adopt $\dot M_2 =0$ and $\dot M = \dot M_1$,we then solve for the time derivative of the binary separation, $\dot{a}$ in equation~\ref{eq:Jdot}. The assumption of a non-accreting companion is an oversimplification, however with our spatial resolution it is difficult to determine the qualitative outcome of accretion versus mass loss from the system.
Also, for systems like Cyg X-1, the fraction of accreted material is expected to be minimal. Estimates from \cite{2003ApJ...583..424G} of the wind mass loss rates from the donor star of Cyg X-1 is found to be  $2.5 \times 10^{-6} M_\odot$/yr. The X-ray luminosity of Cyg X-1 is observed to be $2.4 \times 10^{37}$ erg/s in \cite{2017PASJ...69...52S}. If we assume the X-ray luminosity is due to disk accretion, so that $L= \eta \dot{M}_{\rm{ BH}} c^2$ with $\eta=0.1 $ being the accretion efficiency parameter, we get an accretion rate for the companion $\sim 5 \times 10^{-9} M_\odot$/yr. The fraction of accreted material is then very small, and the assumption of a non-accreting companion more valid.

The value of $\dot{J}$ is a sum of two components. The angular momentum lost from the donor star through winds is $\dot{J}_1 = \dot{M}_1 r_1  v_1$, where $r_1$ and $v_1$ are position and velocity relative to the system center of mass. 
This is the angular momentum content of the gas when it was released from the (non-rotating) stellar surface. Note that $\dot{M}_1$ is negative, so that $\dot{J}_1$ is also negative. This is assuming a spherically symmetric wind, which might not be the case for a tidally deformed star in a close binary \citep{2012A&A...542A..42H}. In practice, the donor star in a close binary is likely to be spun up by tides. But to understand the pure dynamics of the angular momentum in the orbit without including the moment of inertia of the donor star, we use a non-rotating donor star. For a discussion of the effect of donor star spin see Appendix \ref{sec:corot}.

The second component is the binary's torque on the circumbinary wind material. 
The gas that flows within the focusing radii of the companion is compressed in a wake behind $M_2$ making the flow around the binary asymmetric. The torque from the binary is proportional to the gas density, so this asymmetric mass distribution results in a net torque from the binary on the gas. The torque transfers angular momentum from the binary to the gas, effectively dragging the binary. We call this loss of angular momentum $\dot{J}_{\rm drag}$ and we will calculate this contribution in section \ref{sec:Results} from hydrodynamic simulations.

Thus, the total change in angular momentum is $\dot{J} = \dot{J}_{1}  + \dot{J}_{\rm drag}$. To include  $\dot{J}$ in the equation for the evolution of the orbit, we introduce 
\begin{equation}
\gamma_{\mathrm{loss}} = \frac{\dot{J}_{1}  + \dot{J}_{\rm drag}}{ \dot{M}_1} \ \frac{M}{J} = \ \frac{M}{\dot{M}_1} \ \frac{\dot{J}}{J},
\label{eq:gamma_loss}
\end{equation}
so that $\gamma_{\mathrm{\rm loss}}$ represents the specific angular momentum of the ejected gas in units of the specific angular momentum of the binary. Substituting this into equation \ref{eq:Jdot}, we get
\begin{equation}
 \frac{\dot{a}}{a}  = -2 \frac{\dot{M}_1}{M_1}  \left[ 1 - (\gamma_{\mathrm{loss}} +{1 \over 2}) \frac{M_1}{M}  \right],
\label{eq:adot}
\end{equation}
which applies under the restricted conditions of a circular orbit, non-spinning stars, $\dot M_2 =0$, $\dot M = \dot M_1$. 

Notice that since $\dot{M}_1$ is negative, $\dot{a}$ is positive if the term in the square parenthesis is positive, and the binary separation will grow. A critical value $\gamma_{\mathrm{loss, crit}}$, can be defined such that $\dot a = 0$, 
\begin{equation}
\gamma_{\mathrm{loss,crit}}  = q + {1\over 2},
\label{eq:gamma_crit}
\end{equation}
where $q=M_2/M_1$. It is useful to compare this to the dimensionless specific angular momentum of the wind-losing donor star in units of the binary's total angular momentum, $\gamma_{\rm donor}= {M_2 /M_1}$. If $\dot J_{\rm drag}=0$, then the wind angular momentum is unmodified by gravitational interaction with the binary, and $\gamma_{\rm loss}=\gamma_{\rm donor}$. This limiting case of no gravitational drag is often referred to as Jeans mode mass loss, in which case, 
\begin{equation}
\frac{\dot{a}}{a}  = -  \frac{\dot{M_1}}{M},
\label{eq:adot_jeans}
\end{equation}
which shows that as $M_1$ loses mass, the binary separation widens in response to the fractional mass lost.

In general, $\dot J_{\rm drag}$ can have positive or negative sign. In practice, it is typically negative, opposing the orbital motion, and increasing the value of $\gamma_{\rm loss}$. This implies that if $\dot J_{\rm drag}$ is large enough in magnitude, $\gamma_{\rm loss}$ can exceed $\gamma_{\rm loss,crit}$, so that the sign of the orbital evolution reverses from expanding with mass loss to contracting with mass loss. In the following section, we use the theory of BHL flows to show that the wind velocity is an important parameter in determining which behavior results.

\subsection{Expectations from BHL theory}\label{sec:bondi_hoyle}

When the wind from $M_1$ passes by $M_2$, $M_2$'s  gravity  focuses the wind in a wake behind it. The gravitational force of this wake on $M_2$ acts as a dynamical drag, opposing the orbital motion and applying a net torque that changes the orbital angular momentum. 
The BHL approximation describes this gravitational focusing and the development of a wake \citep{1939PCPS...35..405H,1944MNRAS.104..273B,2004NewAR..48..843E}. It has been used to estimate the mass accretion rate onto the companion \citep{1973ApJ...179..585D} and the accompanying torques on the orbit \citep[e.g.][]{2018MNRAS.473..747C,2019A&A...626A..68S}. As we will discuss below, the velocity of the wind is a key parameter, but its interpretation has been complicated in the literature by the accelerating profile of the expanding wind.

The material that is gravitationally captured by the companion passes through the accretion radius  
\beq\label{eq:Ra}
R_a = \frac{2GM_2}{(v_w^2 + v_{\mathrm{orb}}^2)},
\eeq
which 
depends on the relative velocity between the star and the wind.

In what follows, we will assume that the wind velocity $v_{w}$ refers specifically to $v_w(r=a)$, the velocity of the unperturbed spherical wind at the radius of the binary separation -- thus representing the speed of the wind as it passes $M_2$. This choice is motivated by the results of our hydrodynamic simulations, which suggest that the wind velocity as it passes the companion (rather than the wind velocity slope or terminal velocity) plays the strongest role in setting the resulting gas flow around the companion object. 
Finally, we note that the simple expression above initially derived by \citet{1939PCPS...35..405H}, and refined by \citet{1944MNRAS.104..273B}, ignores the gas internal energy, which was added later by \citet{1952MNRAS.112..195B}. The reader is refer to \citet{2004NewAR..48..843E} for an insightful review.

The gravitational drag force can then be estimated as
\begin{align}
F_{\mathrm{drag,BHL}}  & = C_d \pi R_a^2 \rho_w (v_w^2 + v_{\mathrm{orb}}^2)  \nonumber\\
& = C_d \frac{4 \pi (GM_2)^2 \rho_w }{(v_w^2 + v_{\mathrm{orb}}^2)}, 
\end{align}
where $C_d$ is an order unity drag coefficient \citep[e.g.][]{1985MNRAS.217..367S}. Using the continuity equation to substitute the spherical wind density at $r=a$, which gives $\rho_w = \frac{ -\dot{M}_1 }{4 \pi a^2 v_w}$, the drag force can be rewritten as
\begin{equation}
F_{\mathrm{drag,BHL}} = C_d \frac{(GM_2)^2 \dot{M}_1 }{a^2 v_w (v_w^2 + v_{\mathrm{orb}}^2)}.
\label{eq:f_drag}
\end{equation}
This drag force exerts a torque on the companion star around the center of mass. Assuming that the drag force is perpendicular to the orbit, the torque is
\begin{align}
\tau_{\rm BHL} &= r \times F_{\mathrm{drag,BHL}} \nonumber \\
&= {M_1 \over M}    \frac{C_d (GM_2)^2 \dot{M}_1 }{a v_w (v_w^2 + v_{\mathrm{orb}}^2)} 
\end{align}
From this definition of $\tau_{\rm BHL}$ we define a corresponding $\gamma_{\mathrm{drag,BHL}}$ by replacing $\dot{J}_{\mathrm{drag}} = \tau_{\rm BHL}$ in equation \ref{eq:gamma_loss}. In this case, $\gamma_{\mathrm{drag,BHL}}$ can be written as
\begin{align}
\gamma_{\mathrm{drag,BHL}} &=  \frac{M}{\dot{M}_1} \ \frac{\dot{J}_{\mathrm{drag,BHL}}}{J}, \nonumber\\
 &= \frac{C_d G^{3/2}}{a^{3/2}} M^{1/2} \frac{M_2}{v_w (v_w^2 + v_{\mathrm{orb}}^2)}, \nonumber\\
\gamma_{\mathrm{drag,BHL}} &=  \frac{C_d \left( {1 \over q} +1 \right)^{-1}}{ \frac{v_w}{v_{\rm orb} } \left[ \left(\frac{v_w}{v_{\rm orb}} \right)^2  +1 \right] }.
\label{eq:gamma_drag}
\end{align}
The form of the final expression above indicates that BHL theory predicts a dependence of the drag force on two dimensionless parameters: the mass ratio, $q$, and the wind velocity ratio at the orbital separation, $v_w(r=a)/v_{\rm orb}$.

\begin{figure}
    \centering
    \includegraphics[width=0.49\textwidth]{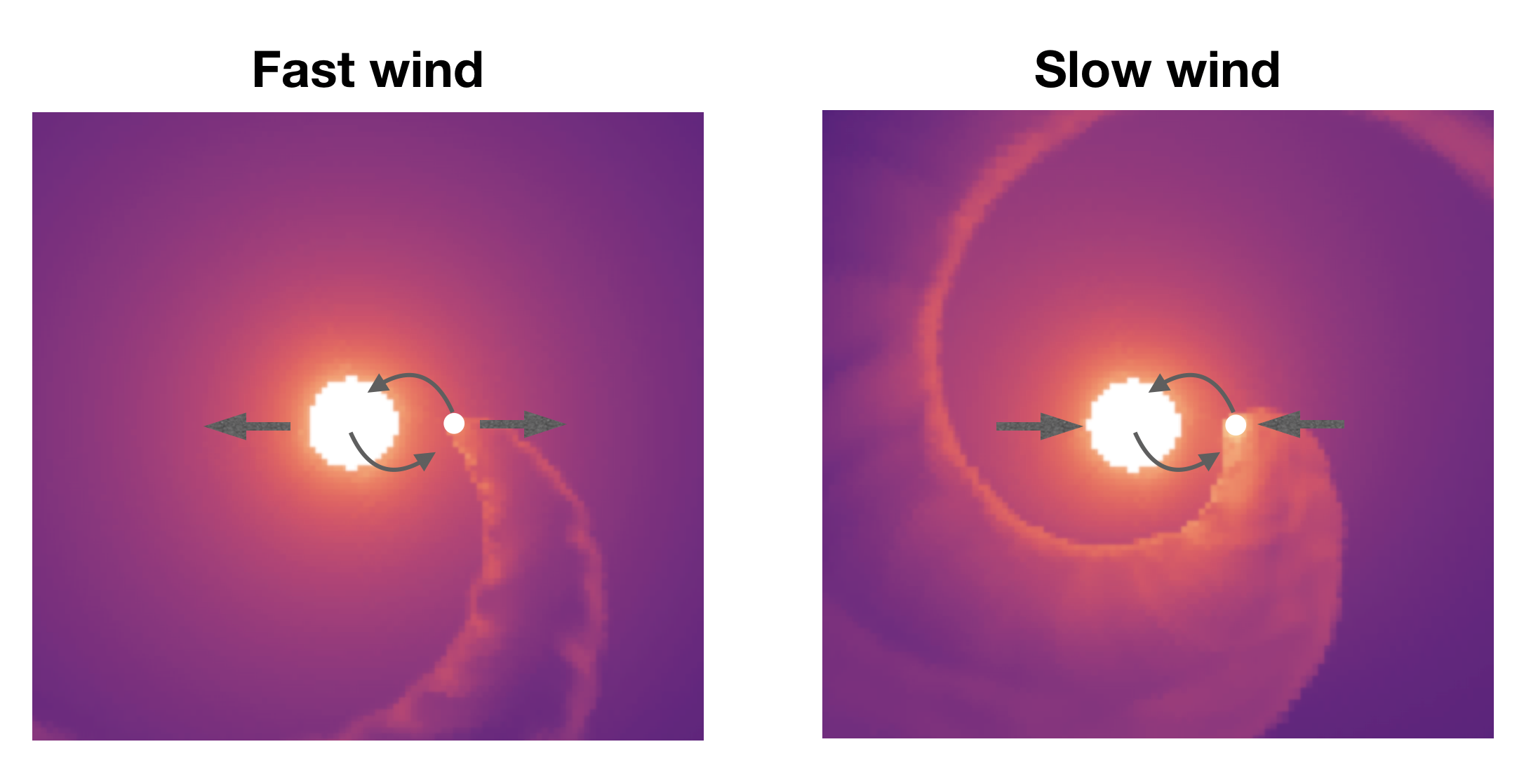}
    \large{BHL theory}
    \includegraphics[width=0.49\textwidth]{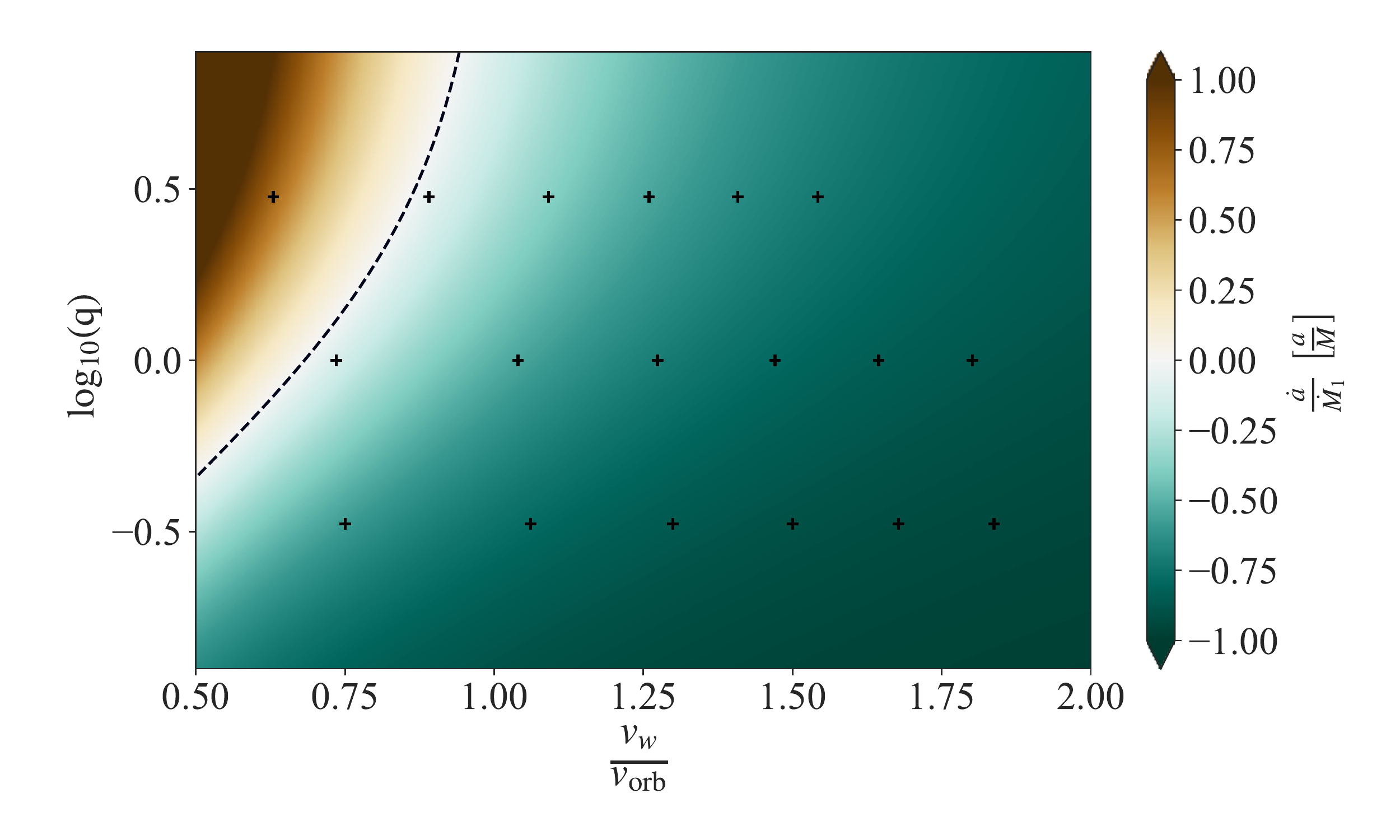}
    \caption{{\it Upper panel}: schematic cartoon showing wind morphology and resulting orbital evolution in cases of higher and lower wind velocity relative to the orbital velocity. {\it Lower panel}: the BHL prediction for the change in orbit per mass lost. The dashed line show where  $\frac{\dot{a}}{\dot{M_1}}\frac{M}{a} = -1$, where the orbit changes from shrinking (brown) to expanding (green) as in the Jeans mass loss case. High velocity winds and low $q$ yield small gravitational drags, and we approach the Jeans mass loss case of $\gamma_{\rm loss} \longrightarrow \gamma_{\rm donor}$. In the opposite limit of higher-mass companions (high $q$) and low velocity winds, $\gamma_{\rm drag}$ can be high enough to drive the orbit inward rather than outward with mass loss. The black crosses indicate the parameter combinations explored in our hydrodynamic simulations discussed in the subsequent sections. }
    \label{fig:adotHL}
\end{figure}

Figure \ref{fig:adotHL} depicts the change in $\dot{a}$ when  $\gamma_{\rm loss} = \gamma_{\rm donor} + \gamma_{\rm drag}$. The crosses indicate the simulations discussed in Section \ref{sec:Results}. When $\frac{\dot{a}}{\dot{M_1}}$ is positive the orbit is shrinking and when $\frac{\dot{a}}{\dot{M_1}}$ is negative the orbit is widening. The dashed black line shows the critical contour where this transition happens. For $\frac{\dot{a}}{\dot{M_1}}\frac{M}{a}=-1$, the orbit changes as in the Jeans mass loss case. Figure \ref{fig:adotHL} demonstrates that this limit is achieved when the wind velocity is high relative to the orbital velocity. In what follows, we use these predictions of BHL theory as a baseline for comparison for our hydrodynamic simulation results.

\section{Numerical setup in Athena++}\label{sec:hydro}

In this section we present the setup we have developed to explore the effects of mass loss in binaries within the \texttt{Athena++} code  \citep{2020ApJS..249....4S}. \texttt{Athena++} is a recently developed Eulerian (magneto)hydrodynamic code descending from \texttt{Athena} \citep{2008ApJS..178..137S}.

\subsection{Fluid equations}
\texttt{Athena++} solves the equations for inviscid hydrodynamics
\begin{eqnarray}
 \partial _t \rho + \nabla \cdot( \rho \mathbf{v}) &  = & 0   \nonumber \\
 \partial _t (\rho \mathbf{v}) +   \nabla \cdot( \rho \mathbf{v v} + P \mathbf{I})   & = &-  \rho \mathbf{a}_{\rm{ext}}  \nonumber \\
 \partial _t E +   \nabla \cdot( [ E + P] \mathbf{v}) & = & - \rho \mathbf{a}_{\rm{ext}} \cdot \mathbf{v}
\label{eq:fluid_eq}
\end{eqnarray}
simultaneously demanding conservation of mass, momentum and energy. Here $\rho$ is the density, $\rho v$ the momentum, $P$ the pressure, $\mathbf{I}$ the three dimensional identity matrix, $E= \epsilon  +  \rho \mathbf{v v}/2 $ is the sum of internal and kinetic energy density and $\mathbf{a}_{\rm ext}$ is an external acceleration that represents the source terms associated with the binary and accelerating wind.

 In our setup, a spherical-polar mesh is centered on the one star that has a wind $M_1$, and we therefore run our simulations in the frame of $M_1$. There are three contributions to  $\mathbf{a}_{\rm ext}$: the gravitational and radiative acceleration from the central star and its wind $a_{1}$, the gravitational acceleration from the companion $a_2$, and the inverted acceleration on $M_1$ in the orbital inertial frame $a_{1i}$, 
\begin{equation}
\mathbf{a}_{\rm{ext}} = \mathbf{a}_{1} + \mathbf{a}_2 +\mathbf{a}_{1i},
\label{eq:sourceterms}
\end{equation}
we do not include the acceleration from the gas on the binary in the integration of the position of $M_1$ and $M_2$. This keeps the binary at the same separation, so we can average the drag from the wind over a longer time for a specific separation and wind velocity.

\subsection{Source Terms}\label{sec:accterms}
The use of acceleration source terms on the hydrodynamics allows us to model the binary motion and the radiative driving of the stellar wind.

\subsubsection{Mass-Losing Star and Wind}
 We begin by discussing the source term for the gravity and radiative acceleration of the mass-losing star $M_1$.  We adopt a simplified version of the \citet{1975ApJ...195..157C} (CAK) approximation for line-driven winds to create the accelerating wind velocity profile for  $M_1$. The radial acceleration term for $M_1$ has a term from the gravity of star $g_{M_1}$, the force on electrons from continuum radiation $g_e$ and the force from Doppler-broadened line absorption $g_L$. For the full derivation of the wind acceleration term and the following wind profile see \cite{1975ApJ...195..157C} equation $20-47$. Here we give a short summery.

In general, $g_L$ arises from the local velocity gradient and Doppler-broadened optical depth. This is shown, for example, in Figure 1 of \citet{1980ApJ...238..196A}, and more thoroughly described in sections 8.6 and 8.7 of \citet{1999isw..book.....L}. However, if we adopt a spherically-symmetric steady state solution for the wind velocity structure, along with the assumption of a homogeneous ionization state (and thus electron-scattering cross section $\sigma_e$), we are able to write a simplified version of the $g_L$ acceleration term that reproduces the wind velocity structure in spherical symmetry. This simplification implies that the driving force has no backreaction depending on the distortion of the wind by the binary gravity, but instead remains spherical. It should thus be regarded as a first-order approximation of the properties of an accelerating wind in a binary system. Under these conditions, we adopt
\begin{align}
\mathbf{a}_1 & = (g_{M_1}  +  g_e + g_L)\hat{{\bf r}},  \nonumber\\
& =   \left( - \frac{GM_1}{r^2} + \frac{GM_1}{r^2}\Gamma_e + {1 \over 1 - \alpha} \frac{GM_1(1 - \Gamma_e)}{r^2} \right)\hat{ {\bf r}},  \nonumber\\
 & =   {\alpha \over 1 - \alpha} \frac{GM_1 (1 - \Gamma_e)}{r^2}\hat{ {\bf r}},
 \label{eq:acc}
\end{align}
where $\Gamma_e = \sigma_e L_\ast/(4 \pi c G M_1) < 1$ is the ratio of the star's electron-scattering Eddington luminosity to its gravity, and $\sigma_e = \sigma_{\rm T} \frac{n_e}{\rho}$ is the electron scattering opacity. We set $\sigma_e = 0.28 cm^2 g^{-1}$ following \cite{1999isw..book.....L} equation $8.93$. The dimensionless parameter $\alpha$ sets the collective strength of the radiation force on Doppler-broadened lines, specifically it is the powerlaw index that relates radiative force to Doppler-broadened optical depth, as described by \citet{1975ApJ...195..157C}, \citet{1980ApJ...238..196A}, and \citet[sections 8.6 and 8.7]{1999isw..book.....L}. We apply $\alpha = 0.73$, which implies that $\alpha/(1-\alpha) \approx 2.7$ \citep{1975ApJ...195..157C}. 

We note that each of $g_{M_1}$, $g_e$, and $g_L$ have $r^{-2}$ scaling, allowing their combination into a composite term. The effect of the $(1-\Gamma_e)$ term is to scale the effective mass of $M_1$, such that we define 
\beq
M_{\rm eff} = M_1 (1-\Gamma_e),
\eeq
and the effective escape velocity of the system is similarly reduced to,
\begin{equation}
v_{\rm eff} = \sqrt{ \frac{2GM_1(1 - \Gamma_e)}{R_1}}. 
\end{equation}
where $R_1$ is the radius of $M_1$, which assumes that the wind is launched from the surface of the star. 

Crucially, the effect of the $g_L$ term is to change the sign of ${\bf a}_1$, such that the composite term is positive (note that ${\bf a}_1\cdot \hat{{\bf r}}>0$). Line driving implies that gas is repelled from $M_1$, rather than retained by it. Under these conditions, there is no hydrostatic atmosphere solution, only wind solutions (except at exactly the Eddington Luminosity).

In isolation from the forces of a companion, we can derive the spherically-symmetric wind velocity profile by integrating
\beq
v_w(r) \frac{dv_w(r)}{dr} \approx {\bf a}_1 \cdot \hat{{\bf r}},
\eeq
from $v_{w}=0$ at $R_1$. We have neglected terms associated with wind gas pressure gradients, an assumption that is valid when the wind is highly supersonic and mostly driven by lines rather than its thermal content \citep[see section 8.7 of][]{1999isw..book.....L}. 
We find,
\begin{equation}
v_w(r) \approx \sqrt{   {\alpha \over 1 - \alpha} 2GM_1 (1 - \Gamma_e ) \left( { 1 \over R_1} - {1  \over r} \right)}. 
\label{eq:CAK}
\end{equation}
From equation~\ref{eq:CAK}, we note that
\begin{equation}
 \sqrt{ {\alpha \over 1 - \alpha}}= {v_\infty \over v_{\mathrm{eff}}}
 \label{eq:vinf_over_veff}
\end{equation}
sets the ratio of the wind velocity at infinity, $v_\infty$,  and the effective escape velocity of the star. This implies $v_\infty \approx  1.64 v_{\mathrm{eff}}$ for $\alpha =0.73$, this ratio is somewhat lower than its observed counterpart \citep[as, for example, shown in Figure 9 of][]{1982ApJ...259..282A}, an effect that is at least partly explained by relaxing the point-source approximation for the radiation field   \citep{1986ApJ...311..701F,1999isw..book.....L}. 

While this simple acceleration term captures some of the crucial features of a line-driven wind, it ignores many elements that are essential in a completely realistic description. Some of these, including multiple scatterings of photons, heating of the stellar photosphere by scattering from the wind, and the instability of solutions in which the radiative force is based on the velocity gradient are discussed in detail in sections 8.10 to 8.13 of \citet{1999isw..book.....L}. The three-dimensional manifestation of instability is a propensity for winds to become clumpy \citep[e.g.][]{2016A&A...589A.102B,2018A&A...620A.150K,2018MNRAS.475.3240E,2020MNRAS.493..447C}. In binary systems with an accreting compact object, an even more crucial effect may be that high-energy irradiation changes the ionization structure of the metals in the wind, reducing the total line equivalent width and line-driving force \citep[e.g.][]{1990ApJ...356..591B,2018A&A...620A.150K}. We discuss the potential impact of some of these simplifications further in Section \ref{sec:discussion}.

\subsubsection{Companion and Reference frame}

 The companion is modeled as a point mass $M_2$ located at the binary separation $a$. The gravitational force from the companion is 
\begin{eqnarray}
  \mathbf{a}_2 & = & - \frac{GM_2}{|  \mathbf{r} -  \mathbf{r_2} |^3} ( \mathbf{r} -  \mathbf{r_2}) \nonumber\\
  & \simeq & - GM_2  | \mathbf{r} -  \mathbf{r_2} |  f_{\rm{spline}} ( | \mathbf{r} -  \mathbf{r_2} | , r_{\rm{soft}})
\end{eqnarray}
 where $\mathbf{r_2}$ is the position of $M_2$ and $r_{\rm{soft}}$ is the softening radius around $M_2$ used in the code. The softening kernel $f_{\rm{spline}} $ is from \cite{1989ApJS...70..419H} and the position of the companion is integrated based on the acceleration from $M_1$ and the additional  acceleration from the non-inertial frame. For a longer description of the how the companion is modelled see \cite{2018ApJ...863....5M} equation 9.

The simulations are done in the frame of the donor star $M_1$. To stay in the frame of $M_1$, the entire system is accelerated by 
\begin{equation}
\mathbf{a}_{1i} = \frac{GM_2}{|\mathbf{r_2}|^3}  \mathbf{r_2}, 
\label{eq:frameacc}
\end{equation}
which reflects the acceleration of $M_1$ by $M_2$ in the inertial frame.

\subsection{Domain and Boundaries}
\label{sec:domain}
The simulations are performed in spherical polar coordinates, originating from the center of $M_1$. The inner-$r$ boundary is set at the Roche Lobe radius \citep{1983ApJ...268..368E} $r_{\rm in}$ of $M_1$, and imposes the wind conditions. For simulations with $q=1$ and $s=3$, we get $R_1/r_{\rm{RL}} = 0.8$, and we will keep this for all simulations. 
We specify the wind on the basis of $\dot M_1$, such that $\rho_{\rm in} = -\dot M_1/(4\pi r_{\rm in}^2 v_w(r)$, with $v_w(r_{\rm in}))$ specified by equation~\eqref{eq:CAK}. This allows us to not specify the mass loss rate and makes the final results scalable to any mass loss rate.

We set the background isothermal sound speed $c_s$ at the inner boundary, and for simulations wit a non barotropic equation of state the non barotropic sound speed is $c_{s'} = c_s / \gamma$. We calculate $c_s$ by assuming black body radiation on the surface of a standardized setup with a $15 M_\odot$ star with radius $8 R_\odot$:
\begin{equation}
c_s = \sqrt{ \frac{  2 k_{\rm B}}{m_{\rm H}} \left(   \frac{c G M_1 \Gamma_e}{R_1^2 \sigma \sigma_e}    \right)^{1/4}      }
\end{equation}
where $k_{\rm B}$ is Boltzmann's constant, $m_{\rm H}$ is the proton mass, $\sigma$ is Stefan-Boltzmann's constant and $\sigma_{e} = \sigma_{T} n_e/\rho = 0.28 \rm cm^2 g^{-1}$ is the electron opacity in accordance with \cite{1999isw..book.....L}. 
Our assumed wind profile is only applicable for super sonic winds. By choosing the inner boundary to be the Roche Lobe radius of $M_1$, we make sure that $v_w(r_in)/c_s(r)>1$ so the wind in all our simulations have supersonic values at the injection radius. 

The outer boundary in the $r$-direction is set up as a diode at $r=10a$, where gas is only allowed to flow out of the grid. 

The $\theta$ and $\phi$ domain covers the full $4\pi$ of solid angle. For the $\theta$-direction boundaries, we employ the ``polar" boundary, while the $\phi$-direction boundary is periodic from $\pi$ to $-\pi$, allowing gas to move through the full sphere.

\subsection{Code Units and Dimensionless Parameters}\label{sec:funky_units}

We run our simulations using  a  set of dimensionless units. This way the measured values of gravitational gas drag can be scaled to fit any set of stellar binary  parameters. The unit of mass is the total mass of the system $M=M_1+M_2=1$. Then the mass of each star is set by the mass ratio $q = {M_2 / M_1}$, so that the mass of the donor star is $M_1 =  (1+q)^{-1} $ and the mass of the companion is $M_2 =  \left( 1+q^{-1}\right)^{-1}$. The unit of length is set to the separation of the binary $a = 1$, and together with the gravitational constant, $G = 1$, the remaining units are set.

The emergent wind from the donor star depends on the donor's radius and luminosity in addition to its mass (equation~\ref{eq:CAK}).  We use the dimensionless Eddington ratio, $\Gamma_e$, to characterize the luminosity. To describe the star's radius relative to the orbital separation, we define,
\beq
s = {a \over R_1}.
\eeq
Together, these properties can be used to calculate the dimensionless parameter $f = v_w (\mathrm{r = a})/v_{\mathrm{orb}}$, or, using equation \ref{eq:CAK},
\begin{equation}
 f = \sqrt{{\alpha \over 1- \alpha} {2(1 - \Gamma_e) \over 1 + q} (s - 1)}.
\label{eq:v_over_vorb}
\end{equation}
Equivalently, this also sets  $f_\infty = v_w (\mathrm{r = \infty})/v_{\mathrm{orb}}$, 
\begin{eqnarray}
 f_\infty &= \sqrt{{\alpha \over 1- \alpha} {2(1 - \Gamma_e) \over (1 + q) }s } = f \sqrt{1 + \frac{1}{s-1}},
\label{eq:vinf_over_vorb}
\end{eqnarray}
the dimensionless wind velocity as $r \rightarrow \infty$. 

In what follows, we explore binary systems of varying dimensionless properties defined by $q$, $s$, $\Gamma_e$, and $f$ in order to assess how each affects a binary's orbital evolution in the presence of a mass-losing star.

\section{Drag forces from winds}\label{sec:Results}
In the following subsections we present three-dimensional simulations following the methodology described in Section \ref{sec:hydro}. A table showing the chosen simulation parameters is given in Table \ref{tab:simtable}. We explore varying $\Gamma_e$ at several fixed values of the mass ratio, $q$, which has the effect of modifying the wind velocity ratio, $f$, equation \eqref{eq:v_over_vorb}.  The resultant torques of all these simulations are compared with the analytical predictions of BHL theory, described in Section~\ref{sec:bondi_hoyle}.

\begin{table}[]
\centering
 \begin{tabular}{cccccccc} 
 \hline
 Name   & $q$    & $\Gamma_e$    & $s$  & $f$   & $f_{\infty}$  & $\gamma_{\rm drag}$    & $\frac{\gamma_{\rm loss}}{\gamma_{\rm donor}}$\\  
 \hline
 A  &   $1$ &    $0.4$ &    $3$ &   $1.8$ &     $2.2$ &     $0.10$ &    $1.10$ \\ 
 B  &   $1$ &    $0.5$ &    $3$ &   $1.6$ &     $2.0$ &     $0.13$ &    $1.13$ \\ 
 C  &   $1$ &    $0.6$ &    $3$ &   $1.5$ &     $1.8$ &     $0.17$ &    $1.17$ \\ 
 D  &   $1$ &    $0.7$ &    $3$ &   $1.3$ &     $1.6$ &     $0.24$ &    $1.24$ \\ 
 E  &   $1$ &    $0.8$ &    $3$ &   $1.0$ &     $1.3$ &     $0.38$ &    $1.38$ \\ 
 F  &   $1$ &    $0.9$ &    $3$ &   $0.7$ &     $0.9$ &     $0.70$ &    $1.70$ \\ 
 \hline
 G  &   $1/3$ &    $0.4$ &    $2.4$ &   $1.8$ &     $2.4$ &     $0.05$ &    $1.15$ \\ 
 H  &   $1/3$ &    $0.5$ &    $2.4$ &   $1.7$ &     $2.2$ &     $0.06$ &    $1.19$ \\ 
 I  &   $1/3$ &    $0.6$ &    $2.4$ &   $1.5$ &     $2.0$ &     $0.09$ &    $1.26$ \\ 
 J  &   $1/3$ &    $0.7$ &    $2.4$ &   $1.3$ &     $1.7$ &     $0.13$ &    $1.38$ \\ 
 K  &   $1/3$ &    $0.8$ &    $2.4$ &   $1.1$ &     $1.4$ &     $0.21$ &    $1.62$ \\ 
 L  &   $1/3$ &    $0.9$ &    $2.4$ &   $0.8$ &     $1.0$ &     $0.41$ &    $2.23$ \\ 
 \hline
 M  &   $3$ &    $0.4$ &    $3.9$ &   $1.5$ &     $1.8$ &     $0.20$ &    $1.07$ \\ 
 N  &   $3$ &    $0.5$ &    $3.9$ &   $1.4$ &     $1.6$ &     $0.25$ &    $1.08$ \\ 
 O  &   $3$ &    $0.6$ &    $3.9$ &   $1.3$ &     $1.5$ &     $0.33$ &    $1.11$ \\ 
 P  &   $3$ &    $0.7$ &    $3.9$ &   $1.1$ &     $1.3$ &     $0.47$ &    $1.16$ \\ 
 Q  &   $3$ &    $0.8$ &    $3.9$ &   $0.9$ &     $1.0$ &     $0.61$ &    $1.20$ \\ 
 R  &   $3$ &    $0.9$ &    $3.9$ &   $0.6$ &     $0.7$ &     $1.02$ &    $2.34$ \\ 
 \hline
\end{tabular}
\caption{Parameters and key dimensionless properties of hydrodynamic simulations. For each model, we list the three defining parameters mass ratio, $q$, Eddington factor, $\Gamma_e$ and stellar compactness, $s$. The following parameters are then derived from these and the  simulations: velocity ratio at the orbital separation, $f$, asymptotic velocity ratio, $f_\infty$, wind angular momentum added by drag,  $\gamma_{\rm drag}$, and total wind angular momentum compared to its initial angular momentum when launched, $\gamma_{\rm loss}/\gamma_{\rm donor}$.   }
\label{tab:simtable}
\end{table}

\subsection{Winds in binaries with varying $\Gamma_e$}\label{sec:Results_f}

\begin{figure}
    \centering
    \includegraphics[width=\linewidth]{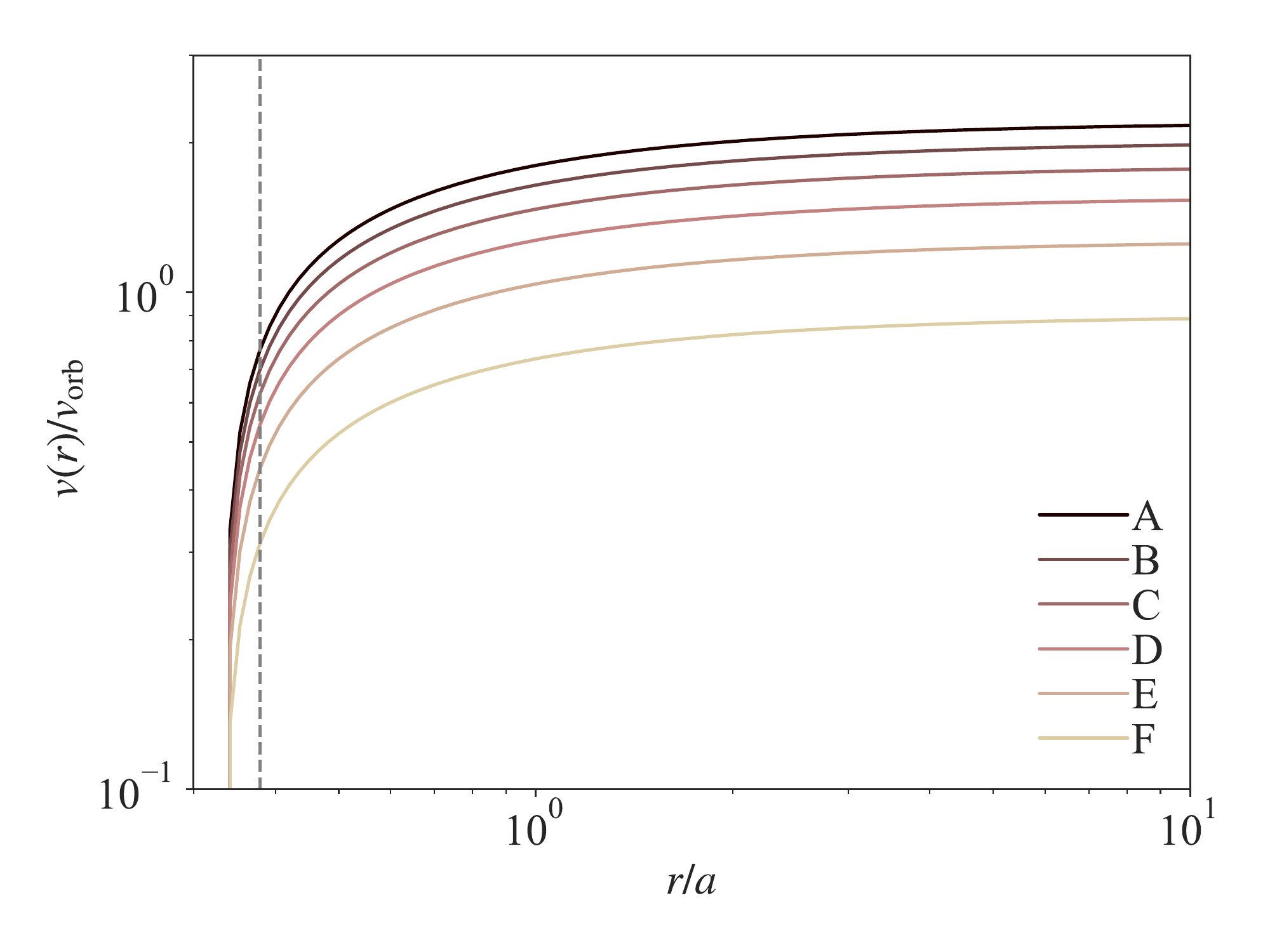}
    \caption{Wind velocity as a function of distance from the center of the donor star for $q = 1$ and $s=3$, and varying $\Gamma_e$. The value of wind velocity compared to orbital velocity is calculated by dividing equation \ref{eq:CAK} with the orbital velocity. The companion is placed at $r/a = 1$, where the wind is still accelerating. The dashed line shows the position of the primary star's Roche Lobe and the position of the inner boundary.}
    \label{fig:wind_profiles_q1}
\end{figure}

\begin{figure*}
    \centering
    \includegraphics[clip,trim=0 8cm 0 10cm, width=1.0\linewidth]{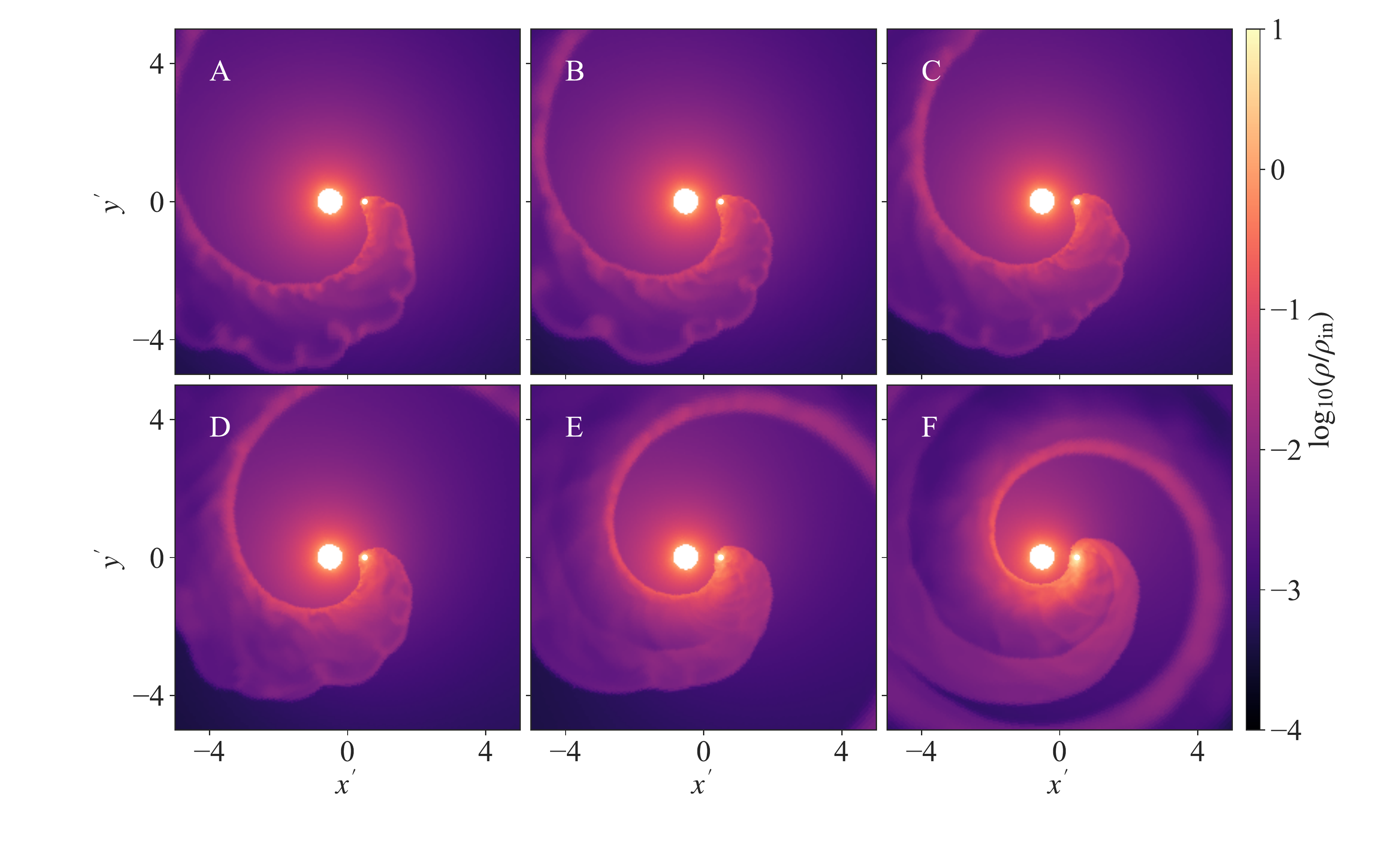}
    \caption{Slice of density divided by density at $r_{\rm in}$ in the x,y-plane  for the six different velocity profiles shown in figure \ref{fig:wind_profiles_q1}. In all cases a tail of focused material is formed behind the companion as it orbits the donor star. But as the value of $\Gamma_e$ increases, the ratio of wind velocity over orbital velocity decreases, and the wake of focused gas forms an continuously tighter and tighter spiral around the binary.}
    \label{fig:q1_s3_rho}
\end{figure*}

In this section we present simulations with varying $\Gamma_e$ between 0.4 and 0.9 for binaries with $q=1$, $s=3$ and gas adiabatic index, $\gamma_{\rm ad} = 4/3$. The choice of $\gamma_{\rm ad}$ was found to have little affect on our the overall results of our simulations, as described in  Appendix \ref{sec:eos}. 
The mesh is constructed with $13 \ (r) \ \times \ 11 \ (\theta)   \ \times \  22 \ (\phi) $ mesh blocks, each of $16^3$ zones on the base level. We then add 2 levels of adaptive mesh refinement (AMR) in the immediate vicinity of the companion mass, $M_2$. For the companion we set  $r_{\rm{soft}}/a = 0.03$. For numerical tests that include variations in $r_{\rm{soft}}$ and spatial resolution, we refer the reader to Appendix \ref{sec:res}. 
We run simulations for two full orbits after steady state is reached, and all derived values are averaged over the two steady orbits.
The radial velocity profiles for the unperturbed wind are plotted  in Figure \ref{fig:wind_profiles_q1} from equation~\eqref{eq:CAK}. As $\Gamma_e$ increases, wind velocities decrease, reducing both $f$ and $f_\infty$. 

These spherical wind profiles are altered in our simulations by the gravity of the companion, which introduces a perturbation in the otherwise radial flow.  The gravity of $M_2$ redirects gas into a converging tail behind it, which expands outwards and forms a spiral around the binary. The resultant modifications of the wind is visualized in Figure \ref{fig:q1_s3_rho}. The various panels show the steady-state density distribution in the orbital plane for six simulations with increasing values of $\Gamma_e$  (A--F in Table \ref{tab:simtable}). Density is shown in units of density at the inner boundary at the donor's Roche lobe, $\rho_{\rm in}$.  All panels in  Figure \ref{fig:q1_s3_rho} are shown in a rotated coordinate system with origin at the center of mass and the binary components along $y' = 0$. The left open circle denotes the location of the donor, $M_1$, while the smaller circle shows the location of  $M_2$, which is responsible for deflecting the wind. 
As  $\Gamma_e$ is increased from simulation A to F, the wind velocity decreases at all radii. This allows allowing for larger wind deviations by the companion object.  The outcome of increasing  $\Gamma_e$ results in a more tightly wound spiral, realized as the wind expands more slowly relative to the orbital motion. A second spiral arm appears in the wake for $\Gamma_e \gtrsim  0.8$. As described by \citet{2018A&A...618A..50S}, the extended inner spiral arm is formed from wind material rotating around the companion in counterclockwise motion that collides with the continuous flux of new wind material from the leading side of the companion object. 
The size and density concentration of the wakes increase with increasing $\Gamma_e$ and decreasing wind velocity.  As we discussed qualitatively  in Section~\ref{sec:bondi_hoyle}, we expect the more massive wake to generate a larger drag force on the binary.

At low wind velocities the gas flow morphology is sometimes discussed as ``wind Roche lobe overflow" \citep{2011ASPC..445..355M}. Winds in this regime expand at low, subsonic radial velocities and gas in the failed wind forms a pressure-supported envelope until it starts mass transferring through $L_1$. Morphologies like this are never observed in our models (even at high $\Gamma_e$ and low relative velocity) because the wind is supersonic and subject to positive acceleration at all radii (as discussed in Section \ref{sec:accterms}, the radiation-driving term is always larger than the local gravitational term), thus no pressure-supported or quasi-hydrostatic solutions exist. We emphasize that this is a qualitative difference in wind morphology that relates to the nature of the wind driving.

\subsection{Winds in binaries with varying $q$}\label{sec:Results_q}
We also investigate wind  interactions in binary systems with varying $q$. We run two extra sets of six simulations (G--L and M--N), varying $\Gamma_e$ within the same range as was done in Section~\ref{sec:Results_f}. For one  of the sets we use  $q=1/3$ while for the other one we use $q=3$. As described in section \ref{sec:domain} we keep the ratio $R_1/r_{\rm{RL}} = 0.8$ the same as in the simulations presented in Section~\ref{sec:Results_f}. Given the different sizes of the Roche lobe radius for the different $q$, we get that $s=2.4$ for $q=1/3$ and $s=3.9$ for $q=3$. 

The resolution around different  companions is kept as close to constant as possible by changing the base mesh dimension to match the new domains,  $10 \ (r) \ \times \ 11 \ (\theta)   \ \times \  22 \ (\phi) $ mesh blocks for simulations with $q=1/3$, and $12 \ (r) \ \times \ 11 \ (\theta)   \ \times \  22 \ (\phi) $ mesh blocks for simulations with $q=3$. All simulations have continue to use $16^3$ zones on the base level and two levels of AMR around the companion.

\begin{figure*}
    \centering
    \includegraphics[clip,trim=0 8cm 0 9.75cm, width=1.0\linewidth]{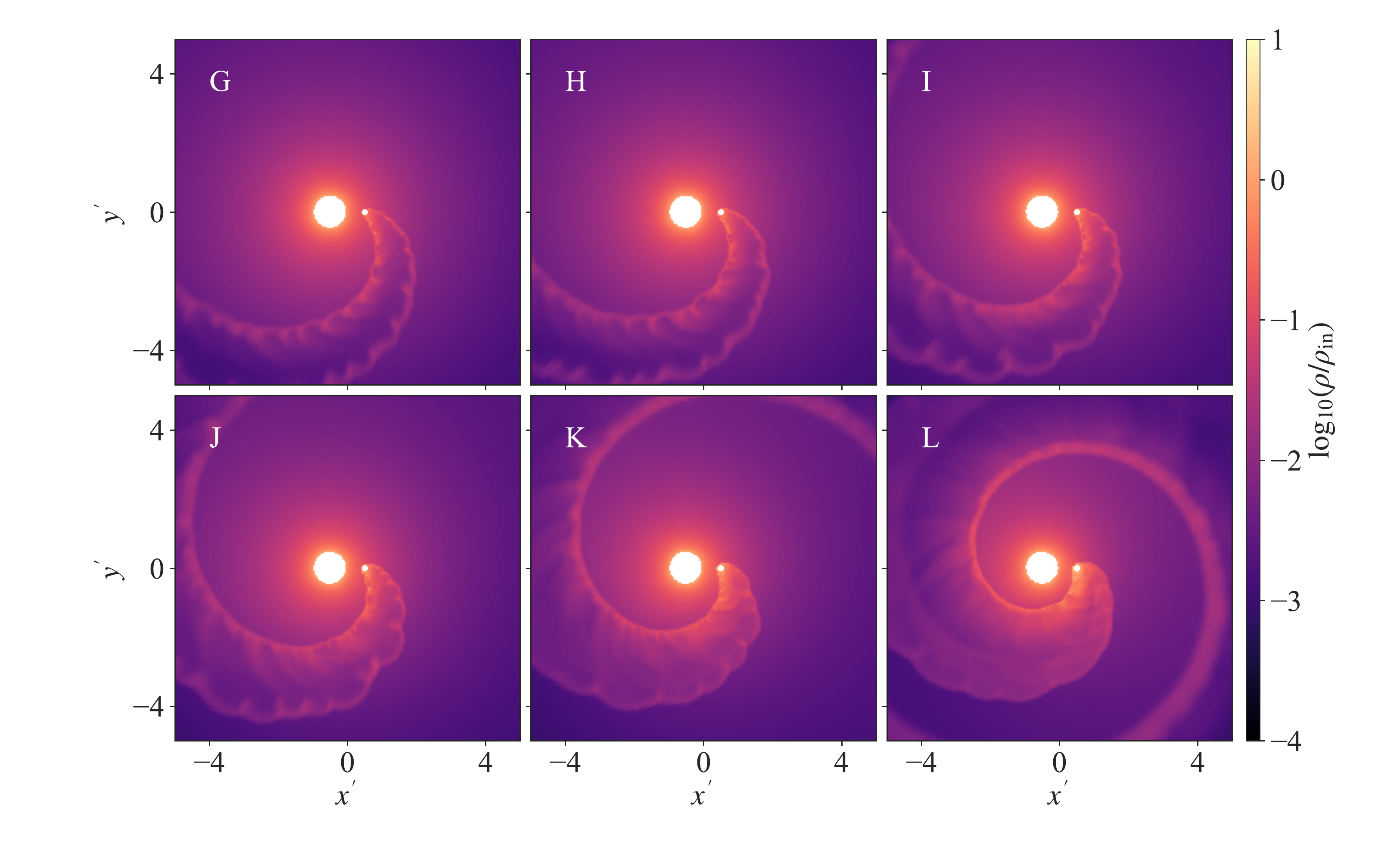}
    \includegraphics[clip,trim=0 8.5cm 0 8cm, width=1.0\linewidth]{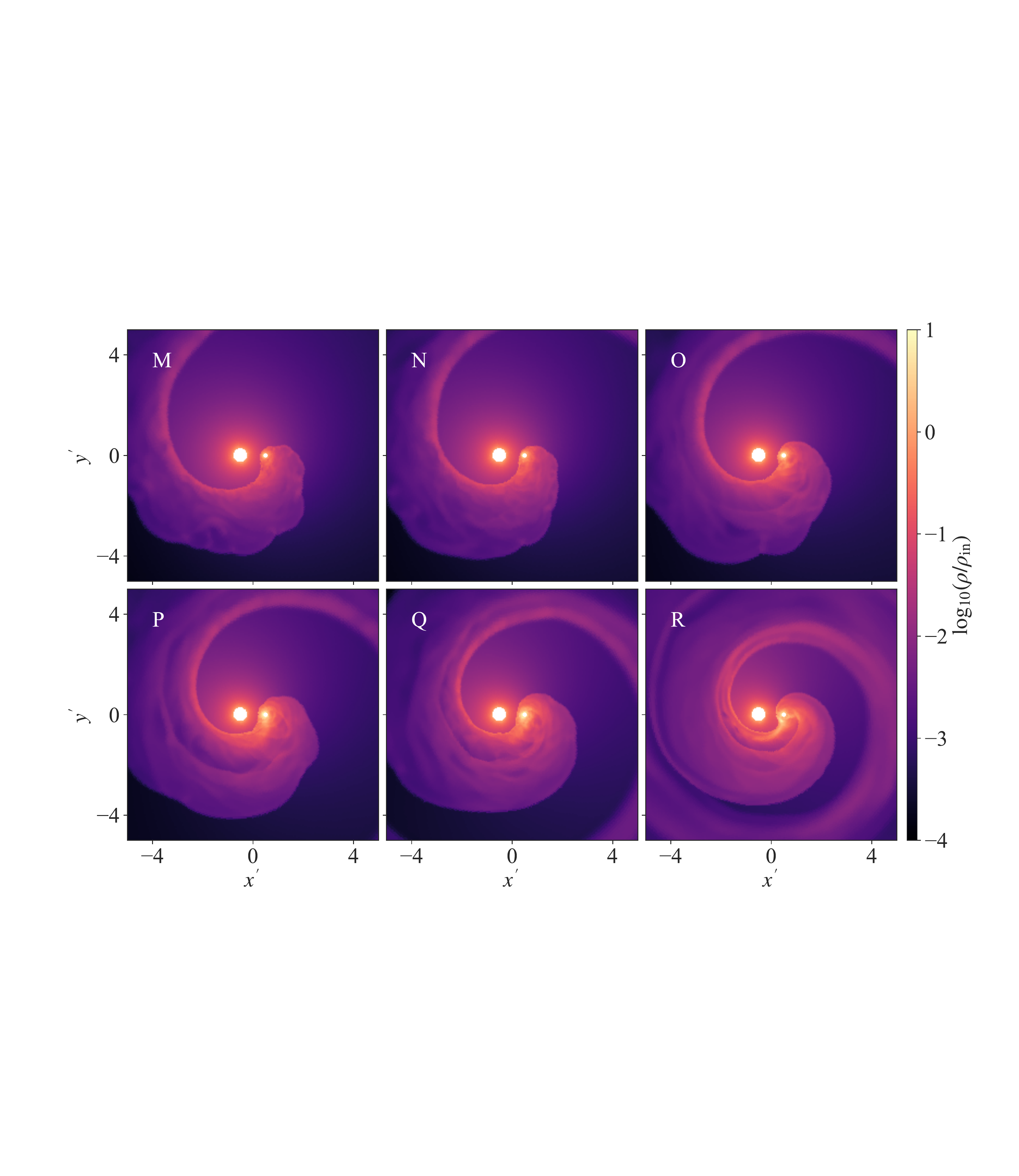}
    \caption{Same as Figure \ref{fig:q1_s3_rho}, but top panel has $q=1/3$ (lighter companion) and bottom panel has $q=3$ (heavier companion). }
    \label{fig:q3_q03_s3_rho}
\end{figure*}

The wind acceleration mechanism employed here depends on  $M_1$. This implies that even for the same value of $\Gamma_e$ a greater radiation force corresponds to a heavier donor star (smaller $q$). This results in different velocity ratios as characterized by $f$ and $f_\infty$, as tabulated in Table \ref{tab:simtable}.

In Figure \ref{fig:q3_q03_s3_rho} we plot the mid-plane density for the two sets of simulations with varying $q$.  Similar behavior with changing $\Gamma_e$ is observed in the simulations with $q=1/3$ and $q=3$ as was discussed above for $q=1$. A noteworthy difference between the $q=1/3$ and $q=3$ cases is the difference in wind density at the companion location relative to $\rho_{\rm in}$. This difference arises because the characteristic scale length for the density gradient is the set by the size of the donor star and not the orbit. The smaller donors (larger $s$) of our $q=3$ cases imply more rapid density fall-off on the scale of the binary. 

The gas focusing behind the companion is weaker for a lighter companion, as we can observe by comparing the panels of Figure \ref{fig:q3_q03_s3_rho}. The spiral structure  is again primarily driven by the value of  $\Gamma_e$, with a more tightly wound spiral for slower winds. But the wake behind $M_2$ is broader and more extended for simulations with $q=3$. A change in $q$ implies  that the gravitational cross section of the companion is thus much bigger for simulations with $q=3$, so more of the wind material is captured, as analytically described by \eqref{eq:Ra}.  

The appearance of a second spiral arm in the wake depends on $q$, with the larger gravitational capture radius of the $q=3$ models, equation \eqref{eq:Ra}, netting material with a broader range of angular momenta relative to $M_2$. As rotating flow around the companion is established at higher velocities, we see the emergence of the the double-spiral structure in more of the $q=3$ models. However, while we expect this trend to be universal, we expect wind capture disks to exist at smaller scales than our companion softening length even when they are not captured in our global models (given a sufficiently compact physical companion). The dynamics of these disks have been recently explored in zoomed-in hydrodynamic simulations by \citet{2013MNRAS.433..295H,2019MNRAS.488.5162X,2019A&A...622A.189E}.

\subsection{Torques and drag forces}\label{sec:Results_torque}

\begin{figure*}[t!]
\centering
\includegraphics[clip,trim=0 8cm 0 9.75cm, width=1.0\linewidth]{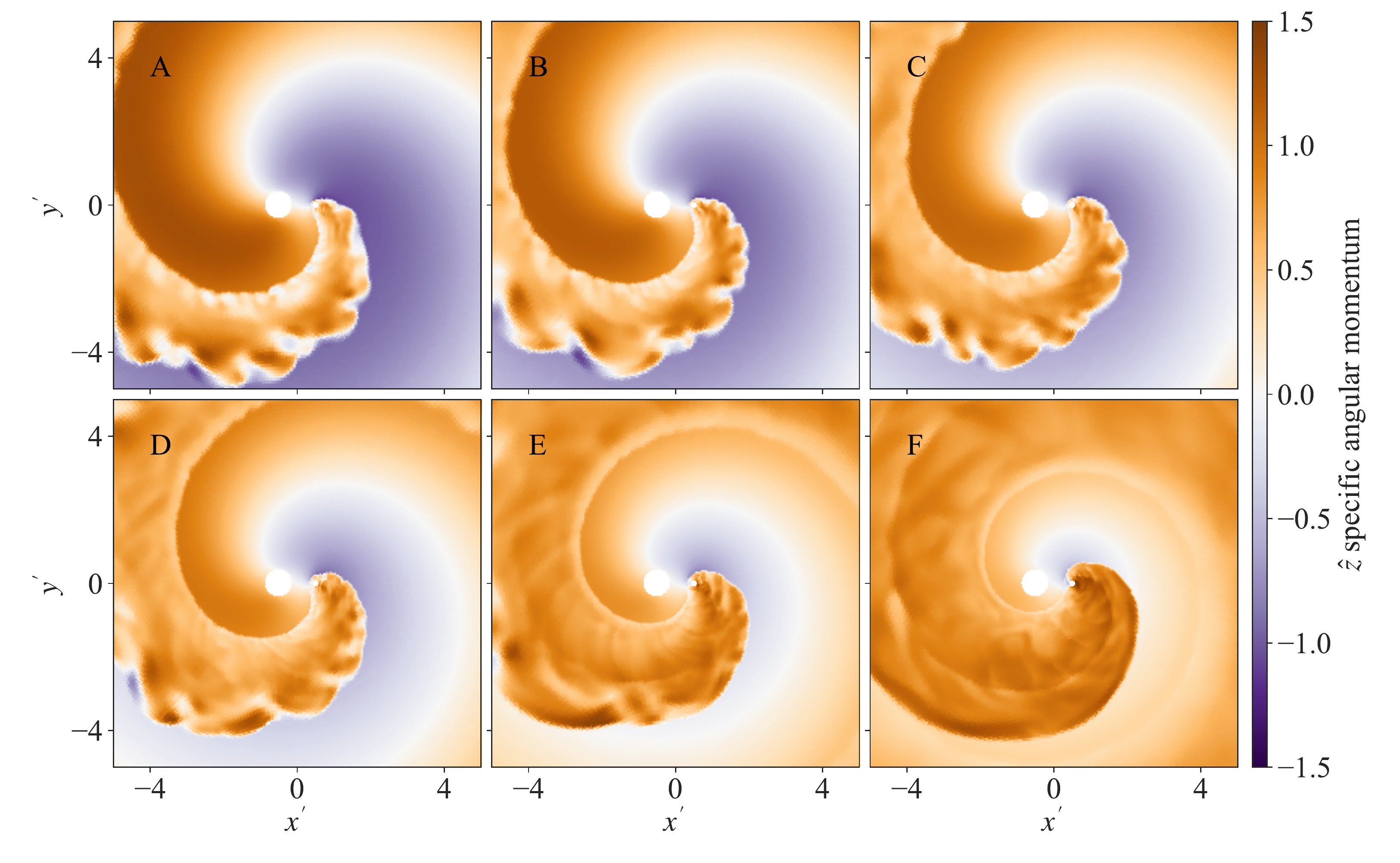}
\caption{Plot of $z$-component of the gas' specific angular momentum around the binary center of mass in a slice in the x,y-plane. The panels are for the same six simulations as shown in Figure \ref{fig:q1_s3_rho}. In our setup, the mass losing star is not rotating. For the slower winds, the specific angular momentum turns predominantly positive, because the gas is effectively torqued and, as a result, the gas receives some of the binary's angular momentum as it expands. }
\label{fig:q1_s3_specLgas}
\end{figure*}

In Figure \ref{fig:q1_s3_specLgas} we plot the $z$-component of the winds' specific angular momentum around the binary center of mass for simulations A--F with $q=1$. In our code units, the specific angular momentum of the donor is $1/4$ when $q=1$. Gas ejected in the direction of the orbital motion of $M_1$  has a positive angular momentum (orange color), while gas ejected in the opposite direction  has negative value (purple color).

In the fast wind case ($\Gamma_e = 0.4$), the distribution of specific angular momentum  has almost equal amounts of gas with positive and negative values, yet the mean is positive (approaching the specific angular momentum of the donor as wind speeds to go infinity). Wind that intersects the wake of the companion gains additional angular momentum through gravitational torques. As $\Gamma_e$ increases and the wind velocity decreases, the range of specific angular momenta within the wind as it is launched narrows (due to the lower wind velocities at the Roche lobe). Additionally, a larger portion of the outflowing wind is captured and torqued to higher angular momentum in the extended wake.

The transfer of angular momentum between the wind and orbit is mediated  by the  gravitational influence from the gas on both $M_1$ and $M_2$. The force felt by each star per unit volume of gas is given by
\begin{equation}
\mathrm{\frac{ \mathbf{F} _{i,gas}}{Vol}  }=  \frac{GM_i \rho}{|  \mathbf{r} -  \mathbf{r_i} |^3} ( \mathbf{r} -  \mathbf{r_i}),
\end{equation}
where $i$ refers to $M_1$ and $M_2$. The resultant force generates a torque on the binary around the center of mass, whose value per cell volume is given by
\begin{equation}
\mathrm{\frac{ \tau _{i,gas}}{Vol}  }= ( \mathbf{r_i} -  \mathbf{r_{\rm com})} \times ( \mathbf{r} -  \mathbf{r_i})  \frac{GM_i \rho}{|  \mathbf{r} -  \mathbf{r_i} |^3} \mathbf{\hat{z}} .
\label{eq:tau_single}
\end{equation}

\begin{figure}
    \centering
    \includegraphics[width=\linewidth]{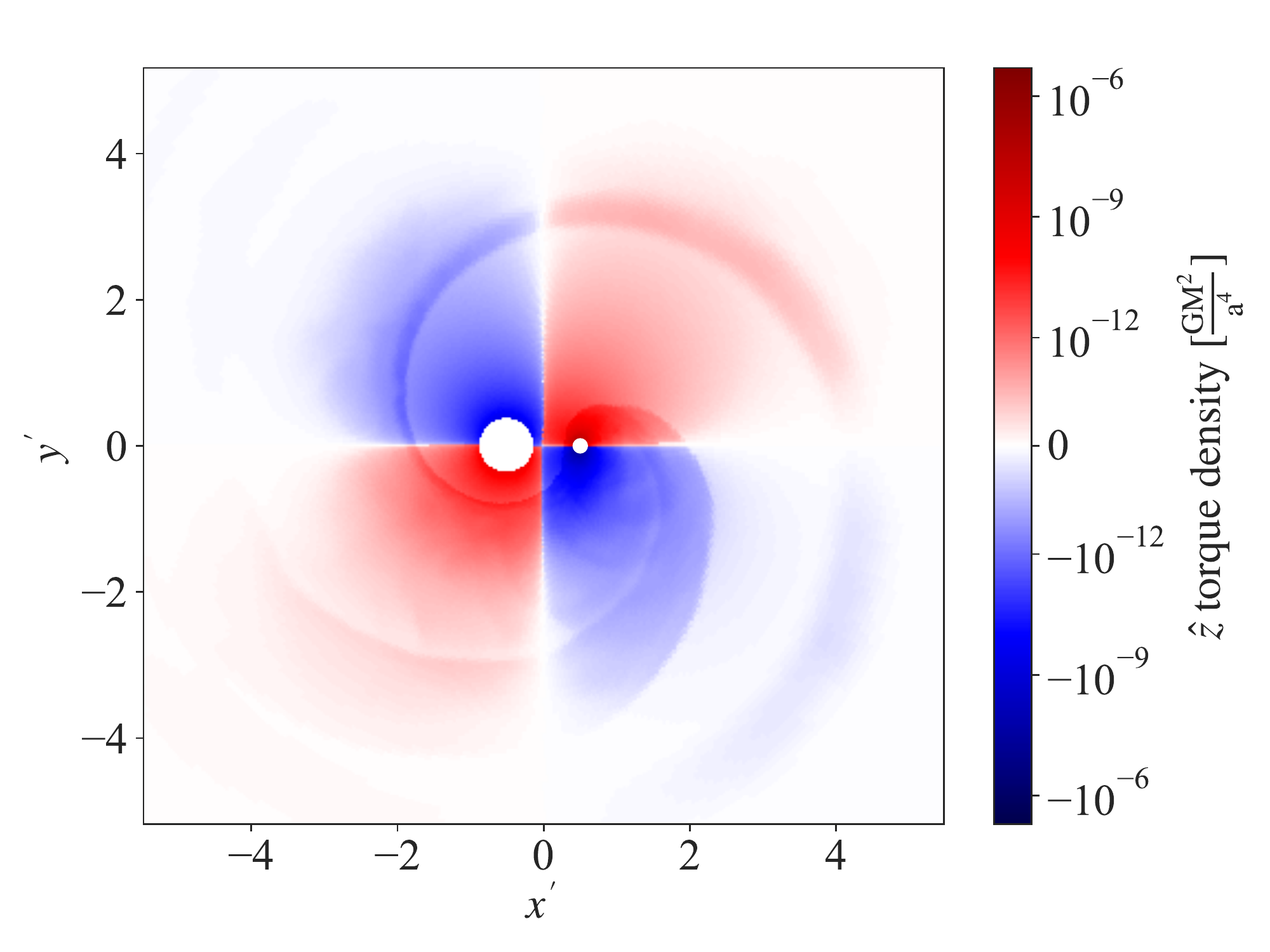}
    \caption{Torque per unit volume calculated with equation \ref{eq:tau_single} and Equation \ref{eq:tau_vol} for the simulation with $q=1$ and $\Gamma_e = 0.9$ ($f=0.75$).}
    \label{fig:tau_dens}
\end{figure}

In Figure \ref{fig:tau_dens} we show the torque per unit volume from Equation \ref{eq:tau_vol} in the binary orbital plane for simulation model F. Regions of the wind contribute both positive and negative torques on the binary. The change in angular momentum of the binary is the sum of the torque on each star
\begin{equation}
\frac{ \dot{J}_{\rm drag}}{ \rm Vol}  =\mathrm{\frac{ \tau _{1,gas}}{Vol}  } +  \mathrm{\frac{ \tau _{2,gas}}{Vol}  }.
\label{eq:tau_vol}
\end{equation}
 The torque on the orbit changes the angular momentum of the binary. Positive drag on $M_1$ pulls the binary forward, while negative drag on $M_2$ pulls it backwards. Figure \ref{fig:tau_dens} shows that the highest torques per volume are located near the binary, in the gas focused behind $M_2$ and in the gas just ejected from $M_1$. Close to $M_1$ the gas is still very symmetric, so the sum of the torque is still minimal. Areas within the high density in the spiral arm also have increased values of the torque. This is not spherically symmetric and will exert a net torque on the binary.

To determine  $\dot{J}_{\mathrm{drag}}$, we compute the sum of the torque from the gas in the entire computational domain as the simulation proceeds and average over three complete orbits. With $\dot{J}_{\mathrm{drag}}$ we then can calculate $\gamma_{\rm drag}$. The angular momentum lost from the donor is calculated making use of equation \ref{eq:adot_jeans} in order to derive $\dot{J}_1$. Equation \ref{eq:gamma_loss} can then be used to calculate $\gamma_{\mathrm{loss}}$, which can then be compared to our analytical predictions (Section \ref{sec:bondi_hoyle}).

\begin{figure}
    \centering
    \includegraphics[width=\linewidth]{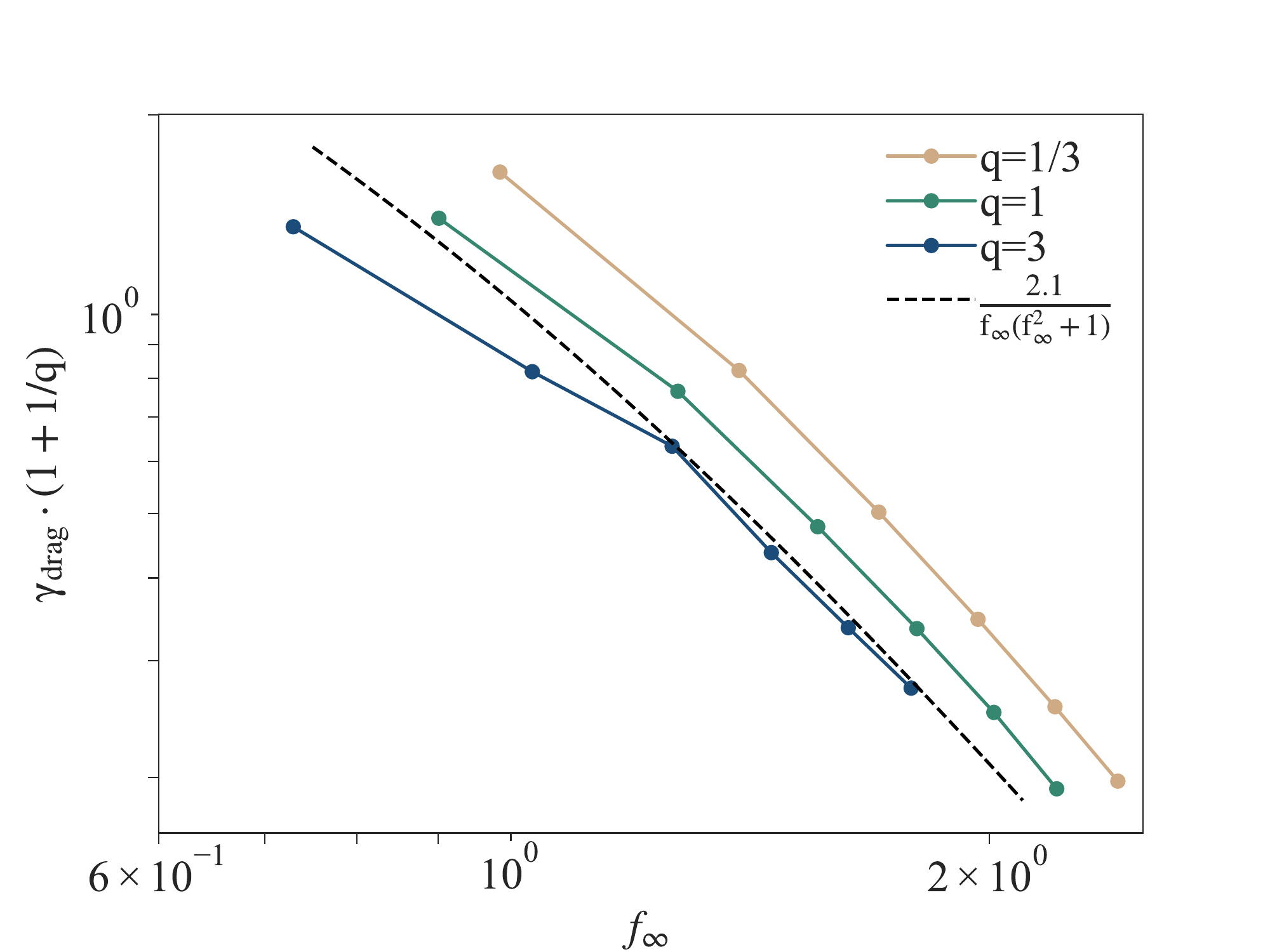}
    \includegraphics[width=\linewidth]{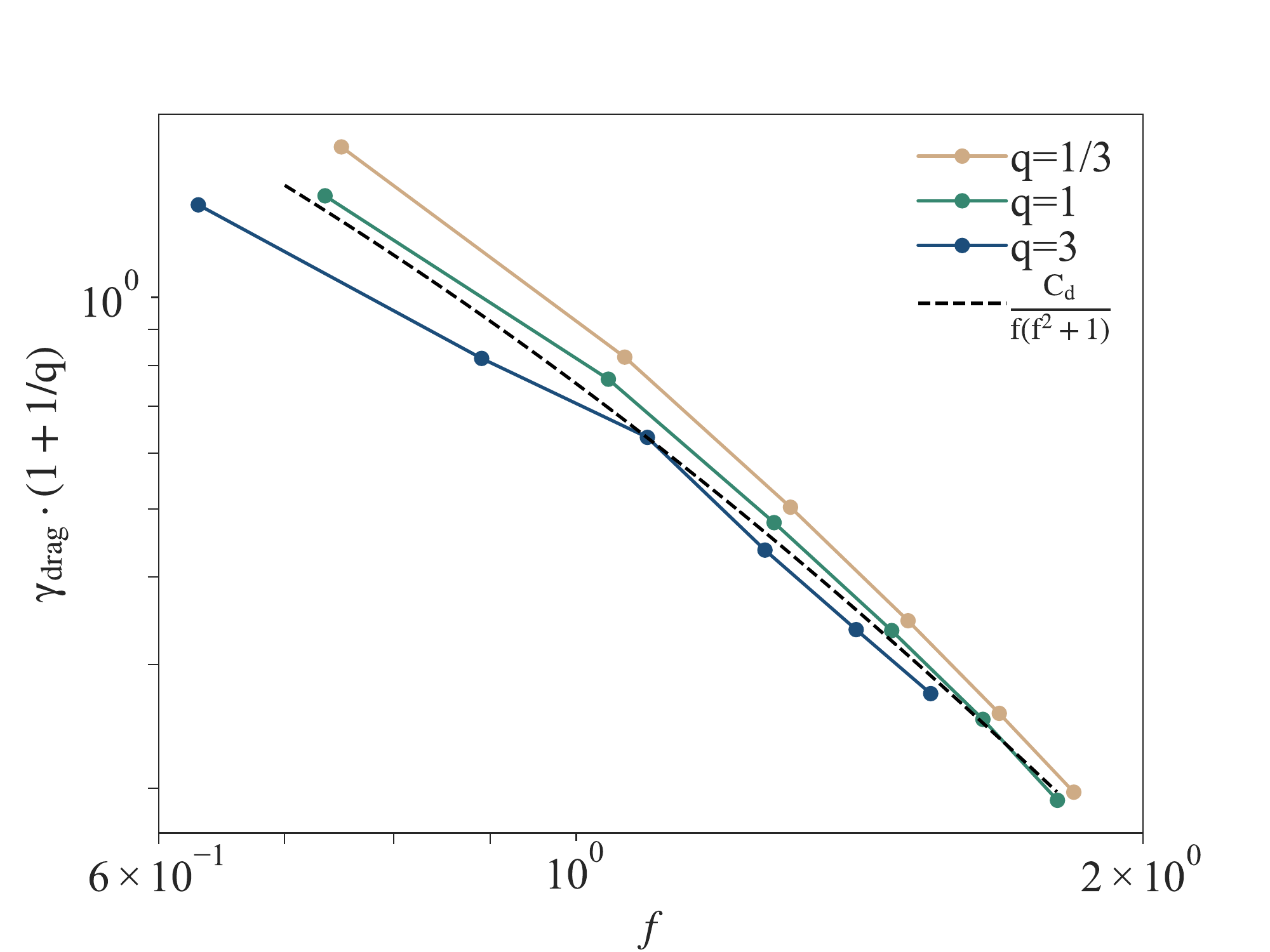}
    \caption{ Here we plot $\gamma_{\mathrm{drag}} \cdot (1+1/q)$ for all simulations A--R. The upper panel shows models plotted as a function of wind asymptotic velocity ratio, $f_\infty$, while the lower panel scales to the velocity ratio at the binary separation, $f$. The dimensionless drag is plotted together with the expected ${1 \over f(f^2+1)}$ dependency from Equation~\eqref{eq:gamma_drag} multiplied with factor $C_{\rm d}= 1.5$. We observe from this comparison that the wind velocity ratio at the orbital separation, $f$, is the primary controlling parameter in the resulting drag force. }
    \label{fig:Ydrag_q1_q3_q03}
\end{figure}

The comparison between our numerical results and the analytical predictions is plotted in Figure \ref{fig:Ydrag_q1_q3_q03} for all simulations with different values of $f$, $f_\infty$ and $q$. The comparison of the upper and lower panels of Figure \ref{fig:Ydrag_q1_q3_q03} demonstrates that $f$, rather than $f_\infty$, is the key dimensionless parameter in determining the dimensionless drag as a function of velocity. This implies that it is not the velocity to which the wind will eventually accelerate that determines its interaction with the binary, but the velocity at distances similar to the binary separation. 

In Figure \ref{fig:Ydrag_q1_q3_q03}, we have scaled the calculated drag term by the mass of the companion, and we have plotted the expected ${C_d \over f(f^2+1)}$ dependency from equation \ref{eq:gamma_drag}. 
By fitting the simulations to our estimate of the drag, we find that a factor of $C_{\rm d} = 1.5$ best describes the data. The change in angular momentum due to wind mass loss can then be effectively described by the following relation
\begin{equation}
\gamma_{\rm loss} = \gamma_{\rm donor} + \gamma_{\rm drag} = q + \frac{1.5 \left( {1 \over q} +1 \right)^{-1} }{ f(f^2+1)},
\label{eq:fit}
\end{equation}
which is the BHL relation augmented with the numerical drag coefficient. 

Broadly speaking, the BHL results of equation \eqref{eq:fit} effectively capture the general trend with $q$ and $f$. However,  differences are most pronounced at lower velocities, $f\lesssim 1.2$. This is related to the fact that the analytical prescription assumes instantaneous acceleration and thus neglects the velocity profile of the wind, which is a  progressively worse assumption over the gravitational capture length scale for lower $f$ values. The velocity slope results in a radially varying density, which is expected to increase the drag \citep{2020ApJ...897..130D,2017ApJ...838...56M}.
We explore and quantify this effect further in Appendix \ref{sec:Results_fesc}.

\begin{figure}
     \centering
     \includegraphics[width=\linewidth]{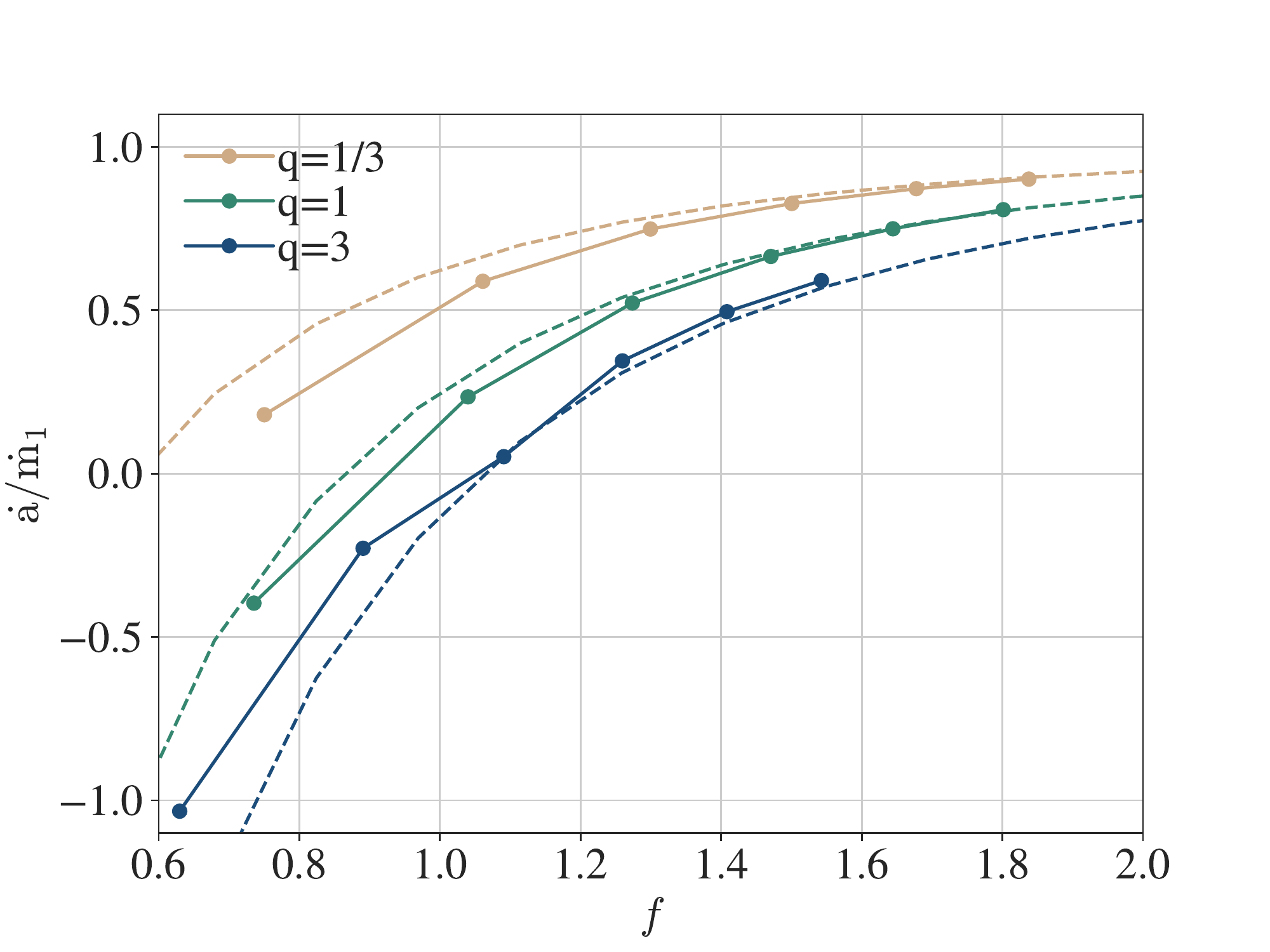}
     \caption{The evolution of the binary's orbit. Shown is the  calculation of $\dot{a}$ from equation \ref{eq:adot} for all simulations with varying $f$ and $q$. Dashed lines show fit from equation \ref{eq:fit}. } 
     \label{fig:adot_q1_q3_q03}
\end{figure}

After calculating $\gamma_{\mathrm{loss}}$ we  use equation \ref{eq:adot} to work out the binary's orbital change.  Figure \ref{fig:adot_q1_q3_q03} shows the value of $\dot{a}$  per unit mass loss. The dashed lines show our best fit to the analytical formalism from equation \ref{eq:fit}. For a fixed $q$, drag forces are most important at low $f$.  As predicted by the analytical formalism (Section \ref{sec:bondi_hoyle}), the drag force needed in order to change the sign of $\dot{a}$ from positive to negative, is smaller for higher values of $q$, and the change in  the binary's orbit increases with decreasing $f$. As can be seen in Figure \ref{fig:adot_q1_q3_q03}, there is a critical value of $f$ for a fixed $q$ when the torque is able to reverse the sign of the orbital evolution, that arises when $\gamma_{\rm drag}=1/2$. 
This critical value, $f_{\rm crit}(q)$,  increases for higher $q$. As such, binary inspiral driven by mass loss takes place when the wind focusing companion is massive and/or the wind is slow.

\section{Discussion}\label{sec:discussion}
In this paper we develop a set of simulations in order to understand the effects of stellar mass loss in  binary systems when one of the stars gradually loses material. We designed the set of  simulations to systematically cover a wide range of mass ratios and wind velocity profiles. In all calculations,  the wind material is focused into a wake behind the companion  and forms a spiral arm around the binary that expands outwards. The exact structure varies with $q$ and $f$.
The velocity of the wind determines  how tightly the spiral arm coils around the binary while the wind velocity profile combined with the mass ratio determine how much gas is effectively focused.  For slow winds and massive companions the nature of the wake is altered. This happens when the circularization radius of the wind material becomes similar to the focusing radius of the companion. In this case, a large-scale centrifugally supported structure  begins to form and  a second spiral arm appears.  For even slower winds a more prominent  disk-like structure  is formed and, as a result, the gas distribution near the companion becomes more symmetric. 

In general, the more asymmetric the distribution of material around the binary is, the stronger is the corresponding torque, which then causes the binary to shrink. In the absence of any gas accumulation, the binary's orbit will naturally expand. As such, slower winds or higher values of $q$ can alter  the  orbit of the binary transforming it from an expanding one into a shrinking one. This behavior can be deduced  from analytical calculations that  make use of the BHL formalism \citep{1939PCPS...35..405H,1944MNRAS.104..273B}, although, as shown by our study, this formalism seems to systematically under predict the resultant drag if one assumes $C_d=1$ (see Section \ref{sec:Results_torque}). The discrepancy is particularly large when the wind is slow compared to the binary's orbital velocity. 

\subsection{Simplifying Assumptions}

We made a number of simplifications in the analysis and models described above.  Some of these are justifiable in wide binaries, but not necessarily in close binaries in which the donor fills a significant fraction of its Roche lobe and the donor’s rotation is tidally locked to the orbit, precisely the regime of slow winds that we are interested in.  We briefly summarise these here.

We generally assumed that the donor was not rotating.  In fact, a tidally synchronised donor will lose extra angular momentum in winds, since each ejected particle carries both the orbital and rotational specific angular momenta.  This is analysed in the model described in Appendix \ref{sec:corot}.

More generally, we ignored the reservoir of angular momentum in the donor’s moment of intertia and rotation.  A tidally synchronised donor will feed back angular momentum into a binary that is widened by winds, exacerbating the widening.  However, since the gyration radius is typically small relative to the orbital radius, this is often a small correction.

A more significant effect is due to ongoing stellar evolution in detached binaries.  The moment of inertia of the donor will typically increase as the donor evolves (until it loses its hydrogen envelope through winds).  Therefore, more angular momentum will be required to keep the star tidally synchronised.  As this angular momentum is taken out of the binary’s orbit, the orbital period will decay, an effect that can mimic the response to interacting winds that we explored.  Many neutron-star HMXBs are observed to have a decreasing period derivative \citep{2015A&A...577A.130F}. This has been attributed to the growing moment inertia of the donors \citep[][see section 5.2 for further discussion]{2000ApJ...541..194L}.

The internal structure of a donor and its moment of inertia  might also be changed by tidal energy deposition, but since tidal dissipation will generally release energy at a much slower rate than the donor’s luminosity for scenarios of interest, this is typically a second-order effect. 

We made a number of simplifying assumptions about the wind profile.  We relax some of these in Appendices \ref{sec:Results_fesc} and \ref{sec:eos}, where we consider winds with different acceleration profiles and clumpy winds.  We also neglect any feedback on the wind profile from the accreting companion, or interaction between two wind fronts if both binary components are losing mass through winds.  

Another significant simplification is the assumption of spherically symmetric winds.  In practice, as the donor star is significantly distorted in close binaries, a spherically symmetric approximation is no longer adequate, and the morphology of the winds will be affected by the donor’s asphericity, including gravity darkening \citep{2012A&A...542A..42H,2020A&A...637A..91E}. This is likely to impact both the predicted evolution of the system and the observationally inferred wind parameters.

We neglected accretion of some of the material in the winds by the companion, but that is likely insignificant for the typical binaries we consider (see Section 2.2).  

We consider the impact of some of these assumptions below.  In section \ref{sec:comparison}, we discuss the validity of our assumptions regarding the equation of state of the wind and compare our results to previous work.  In section \ref{sec:relevance}, we discuss observational constraints from two particular HMXBs, Vela X-1 and Cygnus X-1.

\subsection{Comparison to Previous Studies}
\label{sec:comparison}

When comparing to previous work, it is important for us to highlight some key differences  in approaches, which can be broadly classified in two categories. One relates to the  efficiency of cooling of the gas in the wind interaction region while other one relates to the specific wind acceleration profile assumed in the simulations.

A simple prescription to assess whether the shock interaction region between the stars will be radiative can be obtained using the formalism derived by \citet{2008ApJ...684.1384R}, which compares the cooling length $\lambda_{\rm cool}$ with the separation $a$ of the binary. The shock interaction region will be radiative provided that $(\lambda_{\rm cool}/a)<1$. $\lambda_{\rm cool}$ can be written as \citep{2008ApJ...684.1384R}
\begin{equation}\nonumber
{\lambda_{\rm cool} \over a}=5.5\times 10^{-7}\left({10^{-6}M_\odot{\rm yr^{-1}} \over \dot{M}_{\rm w}}\right)\left({a \over 1{\rm AU}}\right) h({v_{\rm w}}),  
\end{equation}
where 
\begin{eqnarray}
\begin{gathered}
h({v_{\rm w}})=\left[1 + \left(\frac{135 {\rm km\; s^{-1}}}{v_{\rm w}}\right)^{10.7}\right] \times \nonumber \\
\left\{1 - \mathrm{exp} \left[- \left(\frac{v_{\rm w}}{200 {\rm km\; s^{-1}}}\right)^6 \right] \right\} \left(\frac{v_{\rm w}}{100 {\rm km\;s^{-1}}}\right)^{5.2}.
\end{gathered}
\end{eqnarray}

\noindent For $v_{\rm w} = 1000\;{\rm km/s}$, $\dot{M}_{\rm w} = 10^{-6} M_\odot {\rm yr}^{-1}$ and $a = 1 {\rm AU}$,  we get  ${\lambda_{\rm cool} \over a} = 0.09$. This clearly demonstrates that most close  binaries will have $(\lambda_{\rm cool}/a)<1$, thus justifying the use of a more compressible equation of state ($\gamma_{\rm ad}\rightarrow1$ as the gas approaches isothermal).  

In Appendix \ref{sec:eos} we study the role that the equation of state has on the evolution of the binary. That is, we simulate  winds with varying adiabatic index $\gamma_{\rm gas}$, which are used here to broadly simulate the cooling in the interaction region for $\gamma_{\rm gas}<5/3$. By applying this commonly used method we can effectively generalized our results to a broad range of binaries. While the addition of a cooling function makes the results more accurate, the conclusions derived from such an analysis can unfortunately only be applicable to the very specific physical values of a particular  system. The results presented in Appendix \ref{sec:eos} clearly show that the effects of varying $\gamma_{\rm gas}$ are small when compared to those resulting from varying $q$ and $f$. One implication of this conclusion is that the dimensionless character of our simplified models is sufficient for application to real systems for the purposes of estimating wind-driven orbital evolution. 

We also note here that our simulations make use of a simplified wind velocity profile, whereas real winds might follow varying acceleration schemes. This would result in different velocity structures, which are explored in Appendix \ref{sec:Results_fesc}.  Motivated by the results in Appendices \ref{sec:eos} and \ref{sec:Results_fesc}, which show that the key parameters driving the evolution of the binary are $q$ and $f$,  in what follows we present a detailed comparison of our results with those of others. It is important to highlight that although there are clear differences   in approaches taken by the various groups, the results appear broadly consistent, which is encouraging when thinking of constructing generalized prescriptions. 

\begin{figure}
     \centering
     \includegraphics[clip,trim=0 1.2cm 0 1.5cm, width=\linewidth]{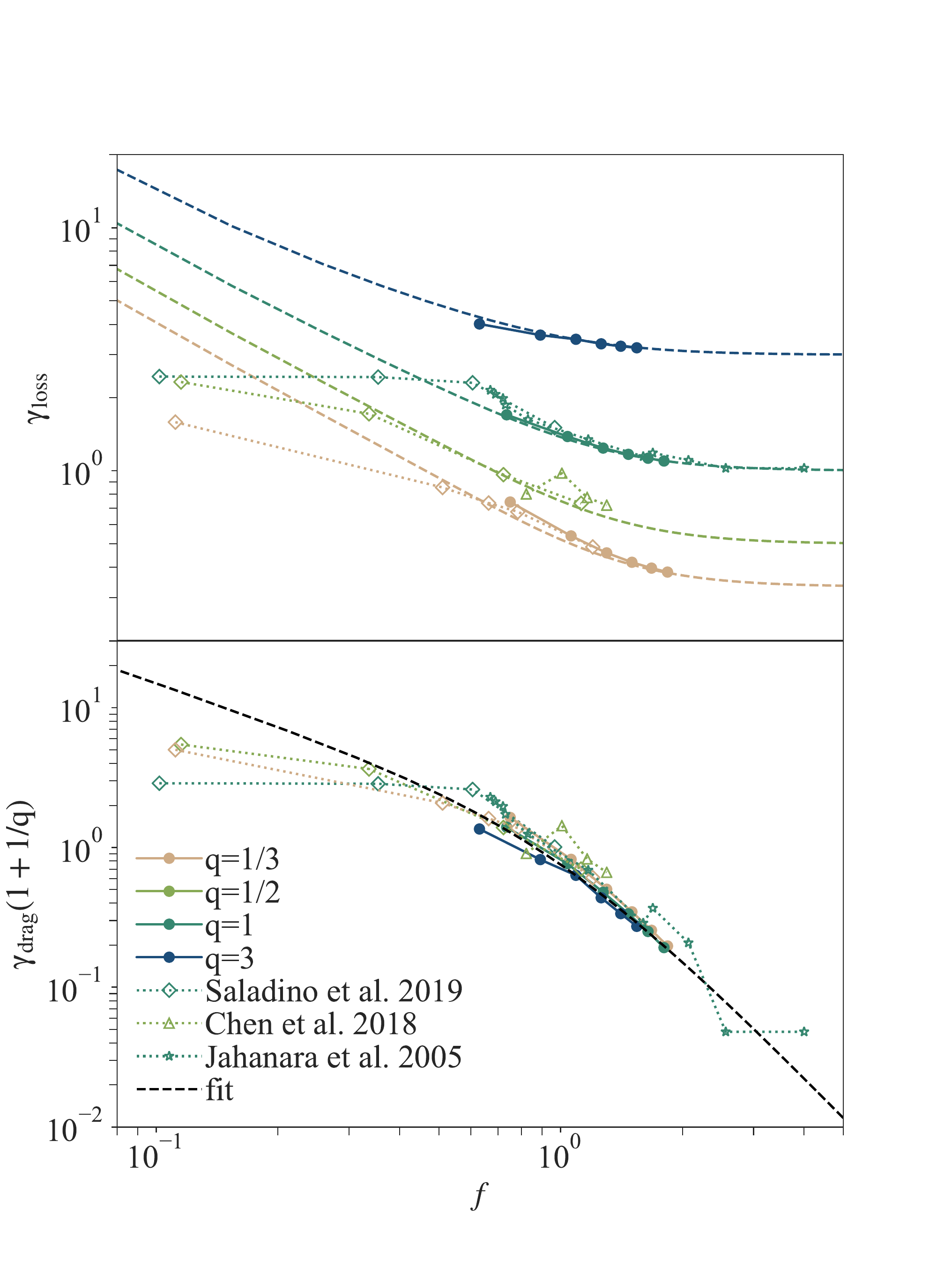}
     \caption{Comparison with previous work. In the top panel we show $\gamma_{\rm loss}$ as a function of $f$ for our simulations in the context of those performed by \citet{2019A&A...626A..68S}, \citet{2018MNRAS.473..747C} and \citet{2005A&A...441..589J}. The data points from \citet{2019A&A...626A..68S} have been translated to values of $\gamma_{\rm loss}$ using equation 12 of \cite{2018A&A...618A..50S}.  \citet{2019A&A...626A..68S} and \citet{2018MNRAS.473..747C} do not explicitly  give  $f$  but instead give the value of the wind velocity at the RL of the donor star, which can be easily translated. \citet{2005A&A...441..589J} give their results for $\gamma_{\rm loss}$ as a function of $f_{\infty}$ rather than $f$. Both \citet{2018MNRAS.473..747C} and \citet{2005A&A...441..589J} run simulations in the corotating frame, which implies  that the wind material is ejected with higher angular momentum due to the spin of the donor star. The simulations performed  here as well as those carried out by  \citet{2019A&A...626A..68S} are not computed  in the corotating frame. In these cases we apply a correction similar to that given in \citet{2019A&A...626A..68S}  in order to derive $\gamma_{\rm loss}$  and $\gamma_{\rm drag}$. In the bottom panel we show the values of the drag component given by $\gamma_{\rm drag} = \gamma_{\rm loss} - \gamma_{\rm donor}$, and multiply with $(1+1/q)$ to get the same quantity as Figure \ref{fig:Ydrag_q1_q3_q03}.} 
     \label{fig:comparison}
\end{figure}

In Figure \ref{fig:comparison} we compare our results  with those from \citet{2019A&A...626A..68S} (diamonds), \citet{2018MNRAS.473..747C} (triangles) and \citet{2005A&A...441..589J} (their radiation-driven cases; stars). \citet{2005A&A...441..589J} run hydrodynamical simulations using  an Eularian setup with $\gamma_{\rm gas } = 5/3$ and drive the wind acceleration by turning off the gravity of the donor star. 
\citet{2018MNRAS.473..747C} run radiation hydrodynamic simulations with the Eulerian code \texttt{ASTROBEAR} using $\gamma_{\rm gas } = 5/3$ with prescriptions for dust formation, gas cooling and pulsations in the mass loss rate, all of which affect the acceleration of the wind material.
\citet{2019A&A...626A..68S} run SPH simulations and similarly to \citet{2005A&A...441..589J},  have a setup with $\gamma_{\rm gas } = 5/3$ and drive the wind acceleration by turning off the gravity of the donor star. They also include terms for cooling and heating due to gas and dust opacity.

In Figure \ref{fig:comparison} we also include the fit given by equation \eqref{eq:fit}, which provides a relatively accurate description of the simulations for $f\gtrsim 1$.  It is clear that for low wind velocities, there are marked differences between the various approaches. As shown by \citet{2019A&A...626A..68S}, the strength  of the torque increases significantly for slower winds and the mass transfer transitions to a Roche lobe overflow as the wind velocity decrease below $\approx 0.9v_{\rm orb}$. In this regime, differences in the acceleration thermodynamics of the wind become relevant, while this is largely not the case at higher velocities.

\subsection{Relevance for observed systems}
\label{sec:relevance}

The orbital period evolution of a binary  system contains critical information about the physics of the binary components and their mutual interactions. In most binaries, the evolution
of the orbital period is too sluggish to be discernible but
among HMXBs this evolution is apparent in a number of cases \citep{1997ApJS..113..367B,2000ApJ...541..194L,2015A&A...577A.130F}.

Since their discovery, HMXBs have been intensively monitored, which has allowed wind accretion models to be tested. Several mechanisms have been invoked to explain the orbital evolution in HMXBs, including  tidal
interaction between the compact accretor and the massive companion as well as wind mass transfer from the massive
component to the compact accretor. The discussion that follows contrasts the evolution of two classical systems: Vela X-1 and Cyg X-1.

Vela X-1 is a well-studied high mass X-ray binary consisting of a $ 2.12 M_\odot$ neutron star in a tight orbit with a $ 26 M_\odot$ donor star ($q=0.08$) with a radius of $29 R_\odot$ \citep{2015A&A...577A.130F}. It has an orbital period of $\approx 9$ days \citep{1995A&A...303..483V}. Here we assume a mass loss rate of $\dot{M} \approx 10^{-6} M_\odot /{\rm yr}$. The terminal wind velocity $v_\infty$ has been estimated using a wide range of methods, where different studies find values between $600$ and $1700$ km/s \citep[for a review see][]{2010A&A...519A..37F}. For this discussion we assume that the wind velocity is equal to the donor star's escape velocity $v_{\rm esc} = 538$ km/s , which gives $f=1.8$. The effective escape velocity from the star includes the effect of luminosity and is likely lower, so this refers to the maximum velocity needed to escape the donor star.

Cyg X-1 has a period of $\approx 5.6$ days, and recent measurements have uncovered that this binary system consists of a $41 M_\odot$  star in orbit with a $21 M_\odot$ black hole ($q=0.5$) at a  separation of $0.24 AU$ \citep{2021Sci...371.1046M}. The size of the donor star was inferred to be $22 R_\odot$, and the mass loss rate of the donor star was estimated by \cite{2003ApJ...583..424G} to be $2.5 \times 10^{-6} M_\odot/$yr.
We again assume that the wind velocity is equal to the donor star's escape velocity, here $v_{\rm esc} = 843$ km/s giving $f=1.7$.

As we have described in this paper, winds can either shrink or expand an orbit depending on $f$ and $q$. The characteristic timescale for winds to alter the orbital period can be written as 
\begin{eqnarray}
\tau_{\rm w} \approx \frac{a}{\dot{a}_{\rm w}} = -{1 \over 2} \frac{M_1}{\dot{M}_1}  \left[ 1 - (\gamma_{\mathrm{loss}} +{1 \over 2}) \frac{M_1}{M}  \right]^{-1},
\end{eqnarray}
which is  the inverse of equation \eqref{eq:adot}. Here $\gamma_{\rm loss}$ is the  fractional change in orbital angular momentum per unit mass loss and is given by equation~\eqref{eq:gamma_loss}. For both systems we expect the orbital period to increase. For Vela X-1 we find $\tau_{\rm w} \approx  2 \times 10^7$ yrs while for Cyg X-1 we infer $\tau_{\rm w} \approx 3 \times 10^7 $ yrs. 

The corresponding period derivative due to winds can then be written as 
\begin{equation}
\frac{\dot{P}}{P} =
\frac{3}{2}\frac{\dot{a}_{\rm w}}{a} - \frac{1}{2} \frac{\dot{M_1}}{M_1 + M_2}.
\label{eq:pdot}
\end{equation}
For Vela X-1 we find $\frac{\dot{P}}{P} = 8 \times 10^{-8} {\rm yr}^{-1}$, and for Cyg X-1 we find $\frac{\dot{P}}{P} = 7 \times 10^{-8} {\rm yr}^{-1}$. 
For both systems we derive increasing periods. This is because the chosen wind velocities are faster than the orbital velocity and these systems have small mass ratios (see e.g. Figure~\ref{fig:adot_q1_q3_q03}).  Observations of Vela X-1 indicate that the period derivative is, however, negative, implying that the system is shrinking \citep{2015A&A...577A.130F}. While this  could be explained if the massive stellar companion had a slow wind, this formalism neglects the effects of tides, which are certainly relevant for these close-in systems \citep{1993ApJ...410..328L,2000ApJ...541..194L}.

Even in the absence of winds, the orbital angular momentum changes because the tidal deformation is phase-lagged with respect to the perturbing tidal forces, which gives rise to a tidal torque that exchanges angular momentum between the orbit and the stellar spin. In general  dissipation leads to the circularization of the orbit  and causes the stellar spins to align and synchronize.  The characteristic timescale for tidal circularization  \citep{2002MNRAS.329..897H} can be written as 
\begin{equation}
\tau_\tau \approx  \frac{2}{21} \frac{R_1 ^{3/2}}{ (GM_1)^{1/2}   q(1+q)^{11/6} \epsilon } (a/R_1)^{21/2}, 
\end{equation}
where $\epsilon = 1.592 \times 10^{-9}   (M_1/M_\odot)^{2.84}$.  For Vela X-1 we derive $\tau_\tau \approx 2 \times 10^5 $ yrs while for Cyg X-1 $\tau_\tau \approx 9 \times 10^3 $ yrs. 
From the above discussion it is evident that the orbital evolution of these systems will be driven by tides until circularization.
Once this equilibrium situation has been achieved the tidal perturbations become stationary in the corotating binary frame and the dissipation ceases. At this stage, the orbit primarily evolves with  the change in the moment of inertia of the massive companion as these changes are mediated to the orbit through the action of tides.

As the star's moment of inertia increases over the course of its evolution, more spin angular momentum is required to keep it tidally synchronised to the orbit.  As this angular momentum flows from the orbit into stellar rotation, the orbit decays \citep{2000ApJ...541..194L}.  This orbital decay is exacerbated by winds which carry away the rotational angular momentum of the mass-losing star, as the surface of a tidally synchronised star may have a  significant additional specific angular momentum relative to its bulk orbital angular momentum.  We consider this latter effect, in the absence of tides, in Appendix \ref{sec:corot}.

An understanding of the roles of winds in altering the orbital evolution in these systems is thus highly sensitive to the exact evolutionary state of the companion and is beyond the scope of this work.  For example, MIST stellar evolution models \citep{2011ApJS..192....3P,2013ApJS..208....4P,2015ApJS..220...15P,2016ApJ...823..102C,2016ApJS..222....8D} for a star with a zero-age main-sequence mass of 43.4 $M_\odot$ star  show that its moment of inertia can grow on a timescale as long as $I_1/\dot{I}_1 \approx 3.6 \times 10^7 $yr during the main sequence or as short as  $I_1/\dot{I}_1 \approx 1.5 \times 10^6$ yr at the end of the main sequence when the supply of hydrogen in the star's core is nearly exhausted. This timescale is comparable to $M_1/\dot{M}_1$ for Cyg X-1 and illustrates the relevance of winds in driving the evolution of binaries with main sequence companions, as in the case of Cyg X-1 \citep{2020AAS...23535504M}. Measuring the change in the orbital period of Cyg X-1 could thus help understand the interplay between winds and tides  and could help uncover the mass loss attributes of the system.  

\section{Summary}
\label{sec:conlusion}
We present a suite of simulations of stellar winds in a binary system using a simplified version of the line driven wind formalism developed by \citet{1975ApJ...195..157C}. We ran simulations with varying wind velocity, mass ratio, wind velocity profile and gas adiabatic index. We used these simulations to study the resultant density structure around the binary and the long term impact on the binary's orbit.
Our key findings are:

\begin{itemize}
    \item The interaction between the wind and the binary creates a spiral pattern in the density distribution. The spiral appears as a result of gravitational focusing from the companion, where a denser wake forms behind the companion (see Figure \ref{fig:q1_s3_rho}). The angle and thickness of the wake depends primarily on the ratio between the wind velocity at the location of the companion and the orbital velocity as well as on the mass ratio.
    
    \item The higher density wake exerts a torque on the binary, which alters its orbit by transporting angular momentum from the stars to the gas (see Figure \ref{fig:tau_dens}). The resultant  torque is largest for slow winds and high companion masses.
    
    \item By comparing the measured torque to the analytic estimate derived in Section~\ref{sec:bondi_hoyle}, we show that the drag measured in the simulations is systemically larger than predicted by theory. We then use the results of the simulations to construct  an approximate formula for the angular momentum loss in equation~\ref{eq:fit}. Comparing our results to previous work shows that the fit works well for wind velocities down to approximately $90\%$ of the orbital velocity, below which differences in the wind acceleration mechanism and differences in gas density across the Bondi radius become crucial to the eventual result (see Figure~\ref{fig:comparison}).
    
    \item Other wind parameters that alter the measured drag are the equation of state of the gas and the  radial velocity slope of the wind within the acceleration region. We discussed these in Appendix \ref{sec:eos} and Appendix \ref{sec:Results_fesc}. More compressible equations of state result in denser and more clumpy density wakes, while a change in the velocity slope induces a momentum gradient across the wake. However, the wind velocity and mass ratio are predominantly responsible for changes in the angular momentum loss and orbital evolution.
    
    \item Finally, we have applied our formalism to the orbital evolution of Vela X-1 and Cygnus X-1 in order to compare the effects of winds and tides in high mass X-ray binaries. We find that tides dominate the orbital evolution of Vela X-1 (as indicated by current measurements), while no period derivative measurements have been reported for Cygnus X-1.
\end{itemize}

Any future study of this problem should refine the treatment of the wind launching mechanism, which depends sensitively on the star's power output. But adding this complexity will make it more challenging to construct generalized prescriptions that  can be implemented in binary population studies covering a wider parameter space. With the simple approximation given in equation \ref{eq:gamma_loss} we hope to begin the refinement of the treatment of angular momentum loss through winds in binary population studies. This will allow us to build a deeper understanding of the orbital  evolution of binaries, which is key when predicting the number of compact binaries currently present in our Galaxy as  well  as providing  direct  tests  of  the formation,  evolution,  mass  transfer  stability  and  merging for all types of binaries.

\section*{Acknowledgements}
We thank A. Vigna-G\'omez for intellectual contributions. We thank the Institute for Theory and Computation at the Center for Astrophysics, Harvard and Smithsonian, for their hospitality while part of this work was completed. We also thank the Kavli Foundation for organizing the Kavli Summer Program in 2017. We acknowledge use of the HPC facility at the University of Copenhagen, funded by a grant from VILLUM FONDEN (project number 16599). The UCSC and NBI team is supported in part by the Heising-Simons Foundation, the Danish National Research Foundation (DNRF132) and the Vera Rubin Presidential Chair for Diversity at UCSC. 
This work was partially supported by the National Science Foundation under Grant No. 1909203. 
TF acknowledges support from the the Swiss National Science Foundation Professorship grant (project number PP00P2 176868).
IM is a recipient of the Australian Research Council Future Fellowship FT190100574.
RWE is supported in part by the National Science Foundation Graduate Research Fellowship Program (Award \#1339067). Any opinions, findings, and conclusions or recommendations expressed in this material are those of the authors and do not necessarily reflect the views of the NSF.

\appendix

\section{Winds with varying velocity slopes}\label{sec:Results_fesc}

\begin{figure}
    \centering
    \includegraphics[width=0.4\linewidth]{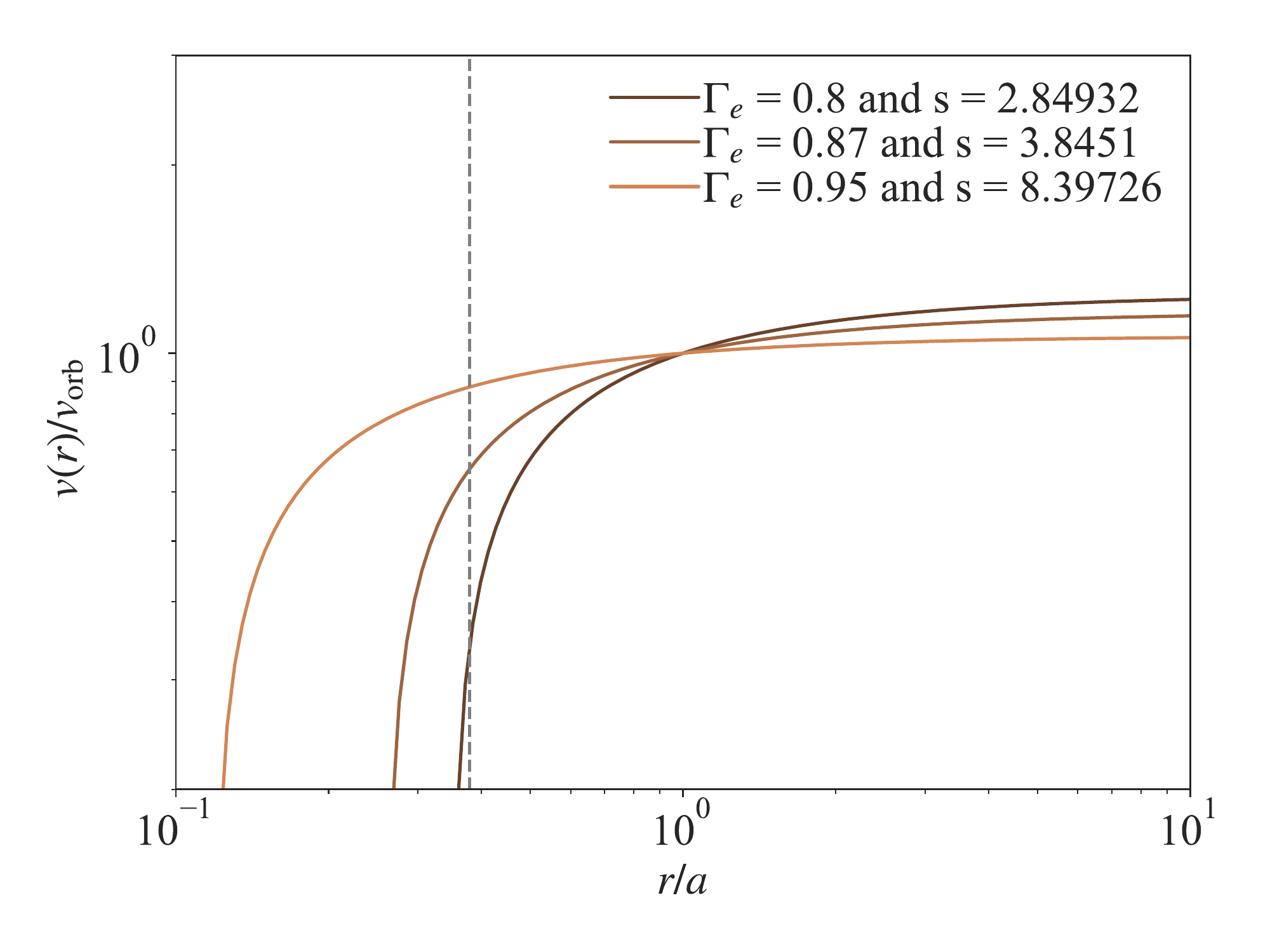}
    \caption{Wind profiles for simulations with f=1. Same as figure \ref{fig:wind_profiles_q1}.}
    \label{fig:wind_profiles_f1}
\end{figure}

In order to investigate the impact of the radially varying wind density we present simulations done with same $f=1$ and $q=1$. These models change the size of the donor star $s$ and its luminosity, $\Gamma_e$, to make the velocity gradient across the position of the companion different. We also calculate drag forces on the binary. The 1D solutions are plotted in figure \ref{fig:wind_profiles_f1}. 

\begin{figure*}
    \centering
    \includegraphics[clip,trim=0 11cm 0 9cm, width=1.0\linewidth]{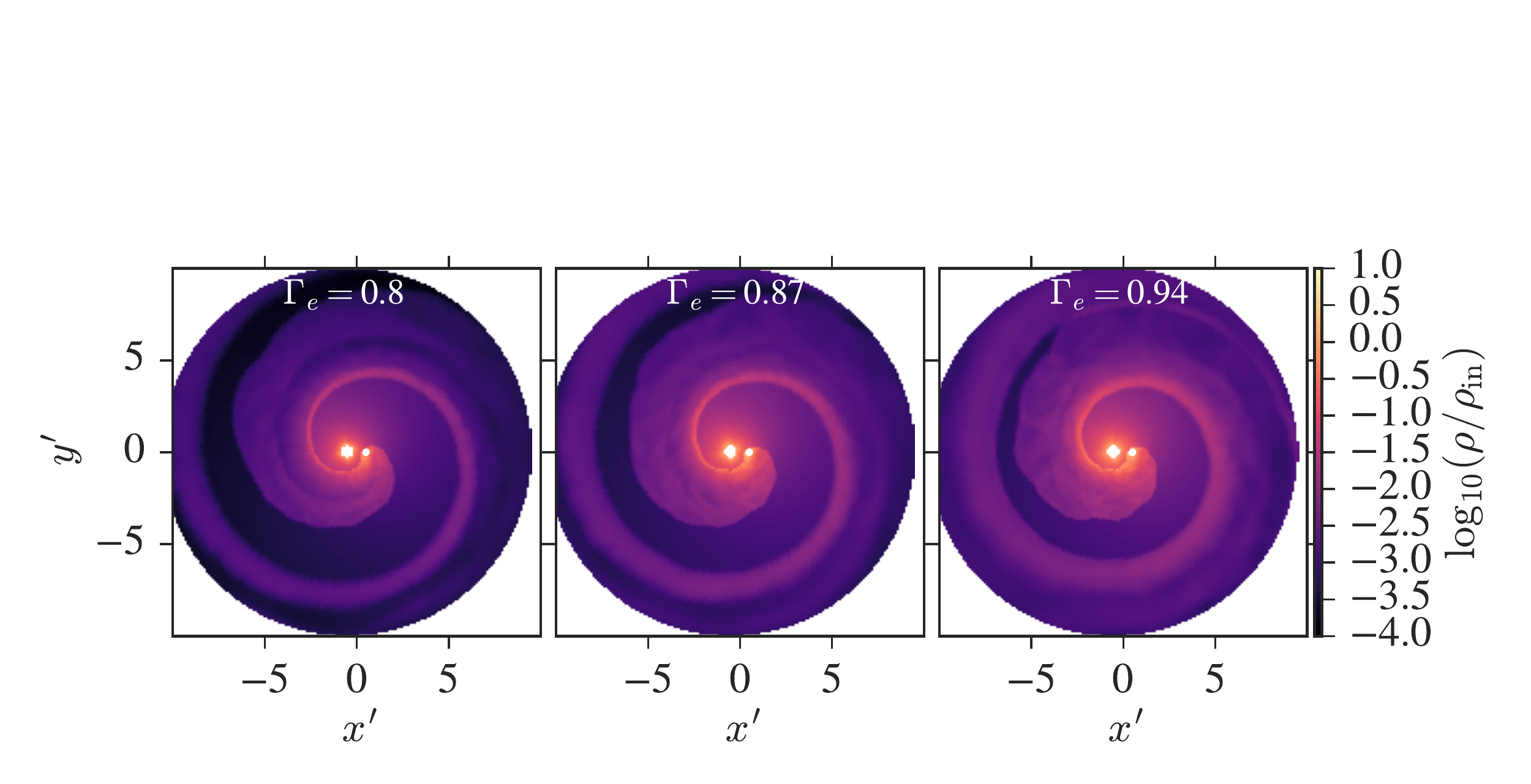}
    \caption{Density slice in the $x,y$-plane across the full computational domain for simulations with velocity profiles from figure \ref{fig:wind_profiles_f1}.}
    \label{fig:f1_rho}
\end{figure*}

We use the same grid setup for these simulations as in section \ref{sec:Results_f}.
The outcomes of the three simulations are visualized in Figure \ref{fig:f1_rho}. The figure shows a slice of density in the $x',y'$-plane across the full computational domain.
The differences in velocity profiles yield different large scale structures, with the flatter profile $\Gamma_e = 0.95$ having a more closed spiral structure. The structure inside the wake is slightly different close to the companion, though the overall shape  is similar.

\begin{figure}
    \centering
    \includegraphics[width=0.5\linewidth]{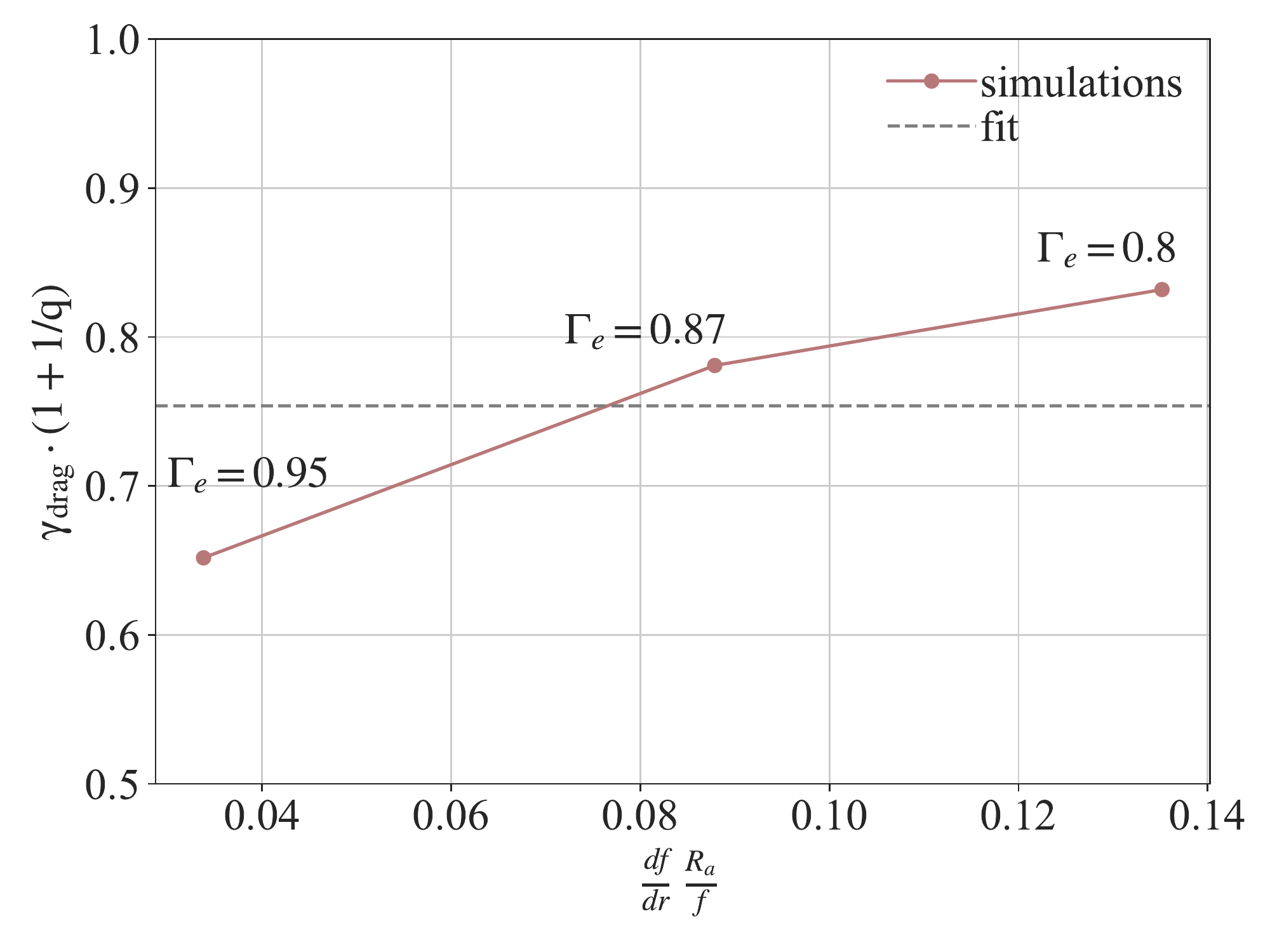}
    \caption{Calculated drag for simulations with the same velocity at the companion, but different velocity slopes. The different velocity slopes give different  momentum gradients across the accretion radius of the companion. The dashed gray line indicate the value expected from equation \ref{eq:fit}. }
    \label{fig:ydrag_f1}
\end{figure}

Figure \ref{fig:ydrag_f1} shows average angular momentum lost from the binary over three periods, same as figure \ref{fig:Ydrag_q1_q3_q03}, but plotted as a function of the velocity gradient $df/dr$ across one $R_a$. For different velocity slopes there is also different gradients in density and in momentum. These gradients are calculated by taking radial derivatives of the equation \ref{eq:v_over_vorb} and multiplying by the $R_a$ to get the momentum difference across the wake. We also divide by the wind velocity $f$, to get fractional change.

Figure \ref{fig:ydrag_f1} shows that the drag is larger for steeper gradients. The steeper gradients also results in larger density variation across the Bondi radius, which  has been found to increase the drag force in wind tunnel simulations \citep{2020ApJ...897..130D,2017ApJ...838...56M}.
And the larger gradient means the spiral tail is less tightly coiled around the binary. We find that these trends can relate to the differences we observe in Figure \ref{fig:Ydrag_q1_q3_q03} in $\gamma_{\rm drag}$ relative to the BHL prediction. Our $q=3$ models have shallower velocity gradients than $q=1$ or $q=1/3$ (due to the smaller value of $s$ and the position of the companion being further down the wind profile). These $q=3$ yield lower normalized drag (relative to the nominal BHL prediction, equation \eqref{eq:fit}), while the $q=1/3$ models exhibit somewhat higher velocity gradient and elevated drag. 

Further study is needed to understand this effect, but these experiments do explain the variation in our simulation results and some of the differences observed between the models of various authors (as discussed in Section \ref{sec:discussion}), especially at low wind velocities. While observations to access complete wind velocity profiles are quite challenging \citep[e.g.][]{2014ARA&A..52..487S}, a simplified approach that applies a characteristic wind velocity profile might provide access to this second-order correction.

\section{Winds from a corotating donor star}
\label{sec:corot}

In this section we will address the change in flow morphology and drag when the donor star is corotating with the orbit. 
We run a new simulations  $\Omega, F$. It has the same parameters as simulation $F$ from table \ref{tab:simtable}, but we run this simulation in a corotating reference frame. This means that two extra source terms are added to the total acceleration from equation \ref{eq:sourceterms}
\begin{equation}
\mathbf{a}_{\rm{ext}} = \mathbf{a}_{1} + \mathbf{a}_2 +\mathbf{a}_{1i} + \mathbf{a}_{\rm{cen}} + \mathbf{a}_{\rm{cor}},
\end{equation}
where $\mathbf{a}_{\rm{cen}} = -\mathbf{\Omega} \times \mathbf{\Omega} \times \mathbf{r}$ is the acceleration due to the Coriolis force and $\mathbf{a}_{\rm{cor}} = - 2\mathbf{\Omega} \times \mathbf{v}$ is the acceleration due to the centrifugal force. 
The wind is still lunched radially without adding any rotation on the surface, so in a corotating reference frame this is equivalent to a donor star with spin synchronized to the orbit.

\begin{figure*}
    \centering
    \includegraphics[clip,trim=1.5cm 5cm 0.5cm 5cm, width=0.9\linewidth]{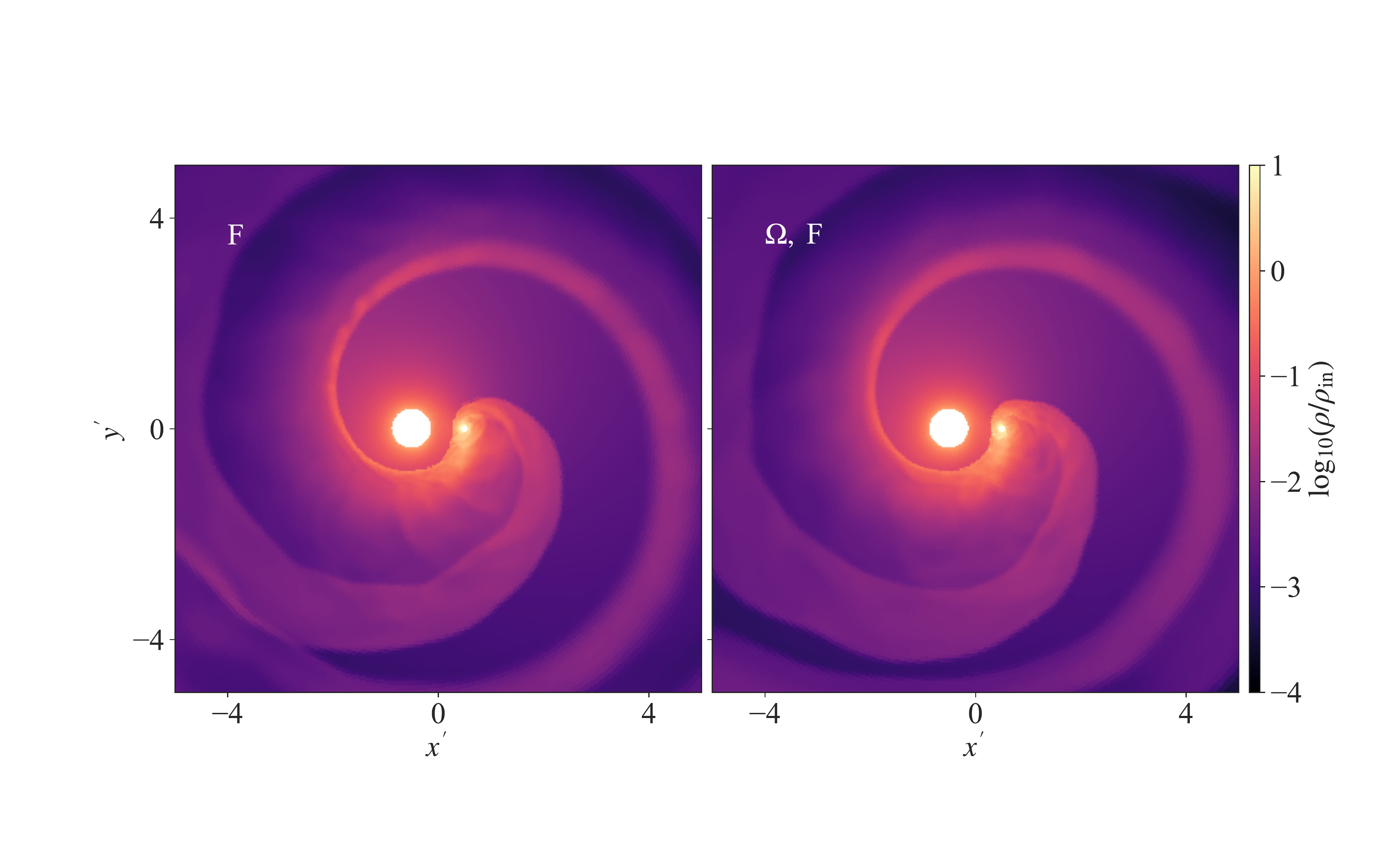}
    \caption{Density in the orbital plane comparing simulation $F$ (left) with new simulations $\Omega, F$ (right) with a corotating donor star.}
    \label{fig:corot_rho}
\end{figure*}

Figure \ref{fig:corot_rho} shows mid-plane density for simulations $F$ and $\Omega, F$. The wake structure is similar in both cases. In the case with rotation the gas has an additional specific  angular momentum $j_{\rm gas} = r_{\rm in}^2 \times \Omega$, but to make a visible difference this extra angular momentum would have to be comparable to the specific angular momentum of the binary, $j_{\rm bin} = \frac{M_1 M_2}{\left(M_1 +M_2 \right)^2} \ a^2 \times \Omega$. The difference in angular momentum thus depends on the mass ratio and the relative size of donor star and the separation of the binary.

Despite the lack of a clearly visible difference in the wake, there is a small difference in the measured drag. For $F$ we get $\gamma_{\rm drag} = 0.69 ^{+0.08}_{-0.1}$ and for $\Omega, F$ we get  $\gamma_{\rm drag} = 0.62 ^{+0.15}_{-0.09}$, meaning that the simulation with a corotating donor may have a smaller value of drag, but the difference is within the measurement uncertainty. \cite{2019A&A...626A..68S} similarly find that the transport of angular momentum from the binary is smaller in simulations with a corotating donor. They also find that the difference is smaller for faster winds. The wind is more symmetric between the two stars, so the measured drag due to asymmetry in outflow is smaller. 
But the gas ejected from a spinning star carries an extra amount of angular momentum matching the stellar surface. The total angular momentum loss should then include this spin angular momentum loss 
\begin{equation}
\gamma_{\rm loss} = \gamma_{\rm donor} + \gamma_{\rm drag} + y_{\rm spin},
\end{equation}
where
\begin{equation}
\gamma_{\rm spin} =   \frac{ 2/3 R_{\rm star}^2  }{M_1 M_2 a^2 }. 
\end{equation}

For the corotating simulation the new total angular momentum loss is $\gamma_{\rm loss} = 1.9$, compared to a non-rotating star where $\gamma_{\rm loss} = 1.7$.  

The difference in measured drag likely depends on the mass ratio and the donor star size compared to the orbit, because it will change the time the orbit is able to torque the emitted wind. Investigations with varying stellar size would be needed to more fully understand how stellar rotation will change the angular momentum transport in winds.

\section{winds with different $\gamma_{\rm ad}$}\label{sec:eos}

\begin{figure*}
    \centering
    \includegraphics[clip,trim=1.5cm 13cm 0.5cm 15.2cm, width=1.0\linewidth]{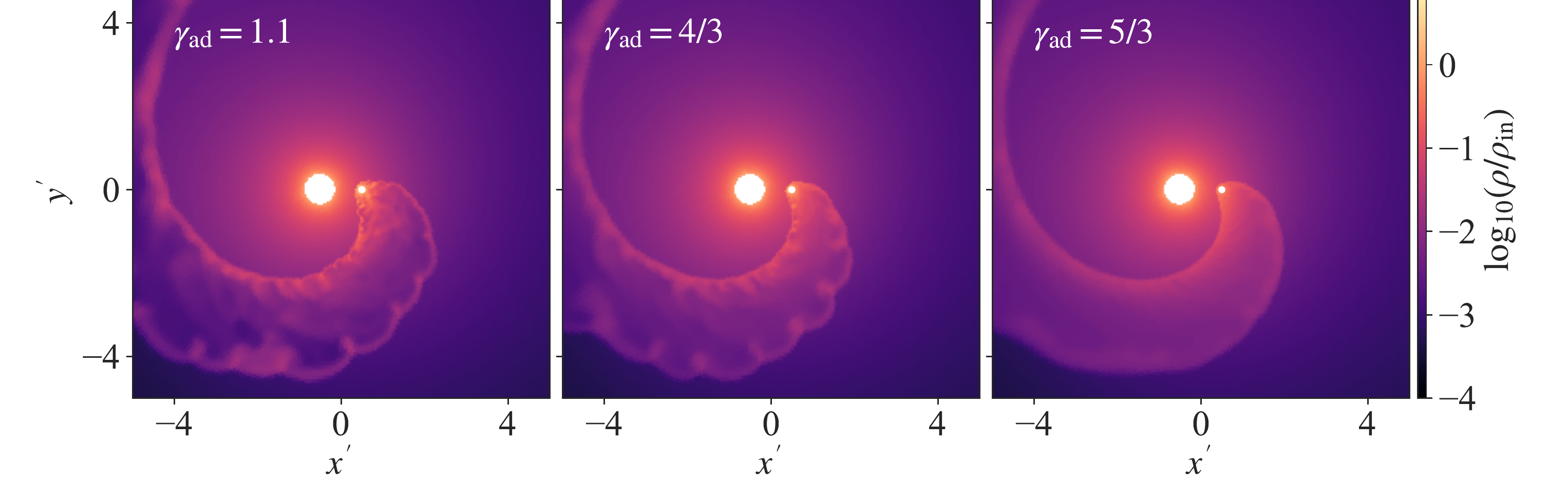}
    \caption{Density in the orbital plane of simulations with $q=1$, $\Gamma_e = 0.4$ and $s=3$ ($f = 1.8$), and varying adiabatic exponents in the equation of state. The shape of the wake is similar for all values $\gamma_{\rm ad}$. The lower value of $\gamma_{\rm ad}$ gives a more compressible and clumpy flow. Values of $\gamma_{\rm drag}$ are $0.48^{+0.07}_{-0.08}$, $0.249^{+0.007}_{-0.009}$ and $0.204 ^{+0.001}_{-0.001}$, respectively, for $\gamma_{\rm ad} =1.1, 4/3$ and  $5/3$ with errors as the $31.8$th to $68.2$th percentile range about the median.}
    \label{fig:eos_rho}
\end{figure*}

The equation of state of the wind is expected to be different for different donor stars. Here we test the simplest variation of the equation of state by using different values for our adiabatic index. In figure \ref{fig:eos_rho} we plot three simulations. All simulations have $q=1$ and $\Gamma_e = 0.4$ ($f=1.8$). The first column has adiabatic index $\gamma_{\rm ad} = 1.1$, the second column has $\gamma_{\rm ad} = 4/3$  and the third column has $\gamma_{\rm ad} = 5/3$.
A smaller adiabatic index allows for more compression for both the fast wind the slow wind case. This leads to more structure in the wake for low adiabatic indices, though the shape of the spiral is the same regardless of the adiabatic index.

We have calculated $\gamma_{\mathrm{drag}}$ following the same method as in section \ref{sec:Results_torque}. The drag is higher for lower $\gamma_{\rm gas}$, where more material is able to gather close to the companion in a symmetric structure. The overall difference is a factor of approximately 2.5 across the range of $\gamma_{\rm ad}$ that we study.  We also include uncertainties taken as the $31.8$th to $68.2$th percentile range about the median. There is a clear pattern from the flow variability seen in this uncertainty, as the lower value of $\gamma_{\rm gas}$ give larger uncertainty. The larger variations with $\gamma = 1.1$ indicate how important gas clumping can be for the measured drag, and suggests further study of the variations with different equations of state.

\section{Numerical Resolution Studies}\label{sec:res}
In this section we investigate the dependence of our results on the size of the companion's softening radius $r_{\rm soft}$ and the levels of AMR in the grid resolution, which affects the spatial resolution near the companion.
The studies presented in this section have $q=1$, $\Gamma_e = 0.9$, and $s=3$.

\begin{figure}
\centering
\includegraphics[clip,trim=1.5cm 11cm 0.5cm 13.2cm, width=1.0\linewidth]{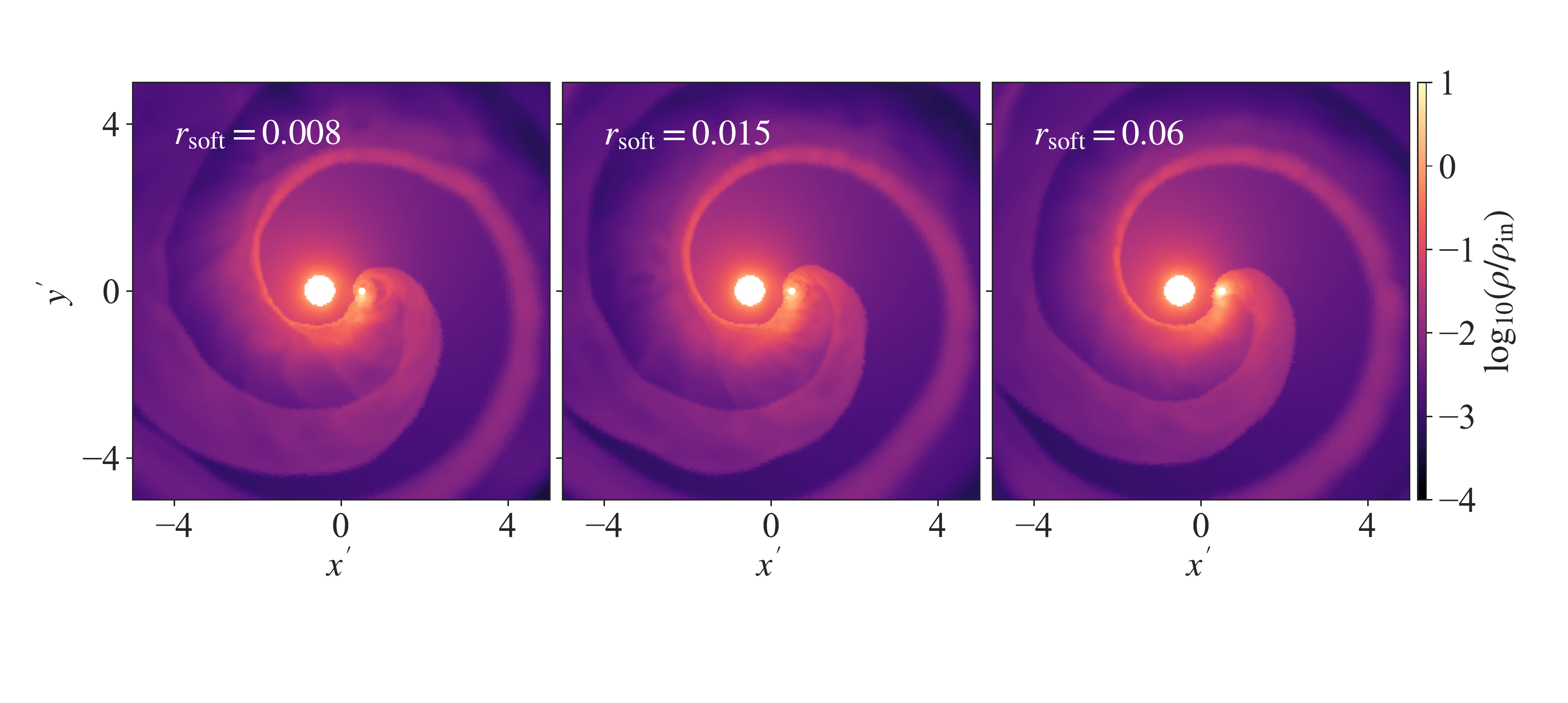}
\captionof{figure}{Density in the orbital plane of simulations with $q=1$, $\Gamma_e = 0.9$ and $s=3$ ($f=0.75$), and varying values of the softening radius $r_{\mathrm{soft}}$ of the companion. The flow structure changes close to the companion for the smallest value of $r_{\mathrm{soft}}$. 
The simulation on the left has $\gamma_{\rm drag} = 1.32^{+0.16}_{-0.17}$, the middle simulation has $\gamma_{\rm drag} = 1.39^{+0.14}_{-0.18}$ and the right simulation has $\gamma_{\rm drag} = 1.45^{+0.13}_{-0.13}$. The simulations in the main text have $r_{\rm soft} = 0.03$ and $\gamma_{\rm drag} = 1.40^{+0.19}_{-0.19}$. The subscripts and superscripts give the $31.8$th to $68.2$th percentile range about the median.}
\label{fig:rsoft_rho}
\end{figure}

We use gravitational softening around the companion to avoid  divergent acceleration, as described in section \ref{sec:accterms}. This is set with $r_{\rm soft}$. In Figure \ref{fig:rsoft_rho} density and $\gamma_{\mathrm{drag}}$ are depicted for variations of $r_{\rm soft}$ where the minimum value is 4 cells in size for our standard resolution (see section \ref{sec:Results}). We ran tests with $r_{\rm soft} = [0.008, 0.015, 0.03, 0.06]$. We choose to run the test simulations for our slowest wind velocities, which have the greatest interaction with the companion. Across this factor of 7.5 in softening radius, we observe only $\pm$5\% difference in $\gamma_{\rm drag}$, assuring us that this is not a primary driver of our measurements.

\begin{figure}
    \centering
    \includegraphics[clip,trim=1.5cm 11cm 0.5cm 13.2cm, width=1.0\linewidth]{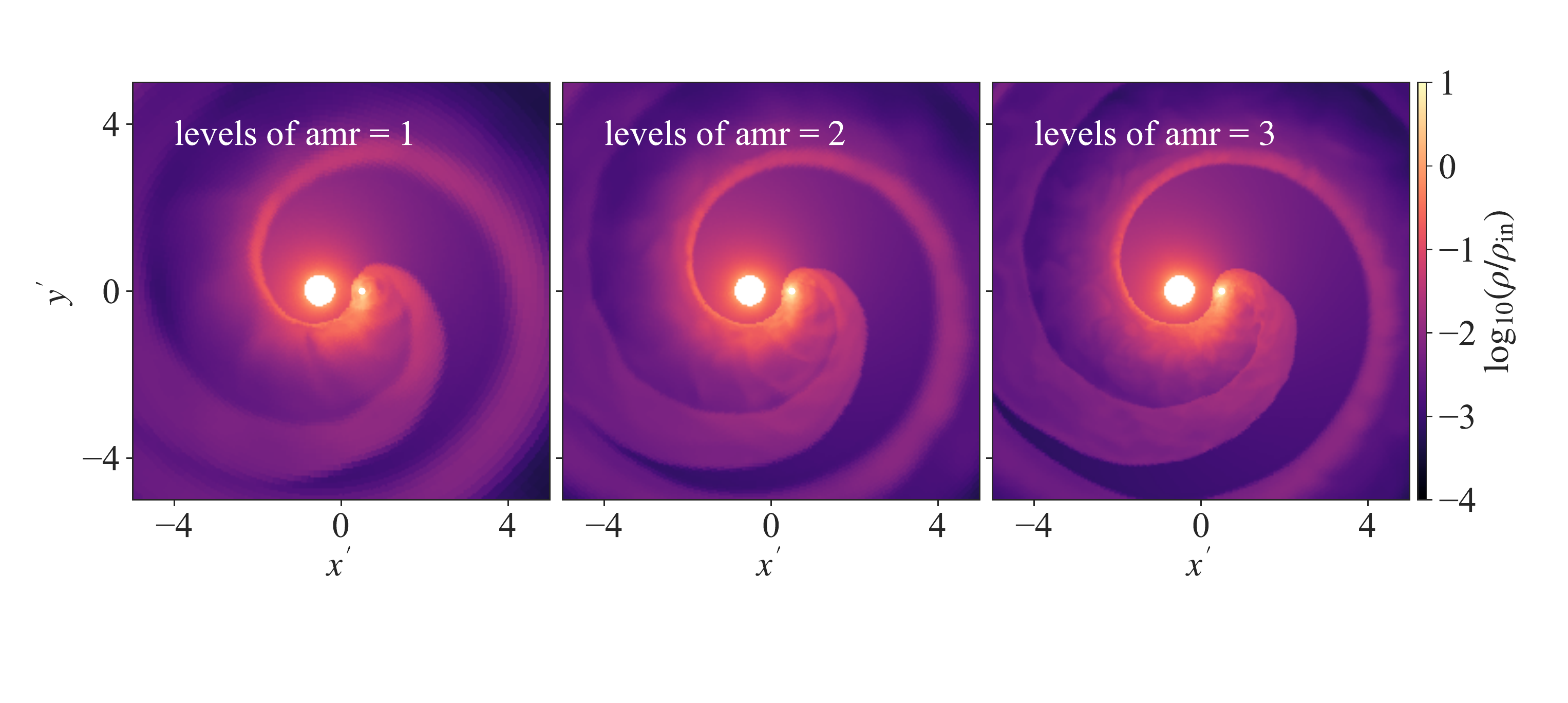}
    \caption{Density in the orbital plane for simulations with $q=1$, $\Gamma_e = 0.9$ and $s=3$ ($f=0.75$), and varying levels of AMR.
    The simulation on the left has $\gamma_{\rm drag} = 1.42^{+0.13}_{-0.18}$, the middle simulation has $\gamma_{\rm drag} = 1.40^{+0.19}_{-0.19}$ and the right simulation has $\gamma_{\rm drag} = 1.41^{+0.17}_{-0.14}$. The subscripts and superscripts again give the $31.8$th to $68.2$th percentile range about the median.}
    \label{fig:amr_rho}
\end{figure}

The spatial resolution set by the level of AMR around the companion object can also change the distribution of the gas. We test 1, 2, and 3 levels of AMR around the companion. The result is shown in figure \ref{fig:amr_rho}. The large scale flow structure is similar in all cases, so the average measured $\gamma_{\rm drag}$ is also similar.  Only the finer structure of the flow is clearer for higher levels of AMR. We use 2 levels of AMR for our production runs.

\newpage
\bibliographystyle{aasjournal}
\begin{scriptsize}
\bibliography{export-bibtex}

\begin{thebibliography}{}
\expandafter\ifx\csname natexlab\endcsname\relax\def\natexlab#1{#1}\fi
\providecommand{\url}[1]{\href{#1}{#1}}

\bibitem[{{Abbott}(1982)}]{1982ApJ...259..282A}
{Abbott}, D.~C. 1982, \apj, 259, 282

\bibitem[{{Abbott} {et~al.}(1980){Abbott}, {Bieging}, {Churchwell}, \&
  {Cassinelli}}]{1980ApJ...238..196A}
{Abbott}, D.~C., {Bieging}, J.~H., {Churchwell}, E., \& {Cassinelli}, J.~P.
  1980, \apj, 238, 196

\bibitem[{{Bildsten} {et~al.}(1997){Bildsten}, {Chakrabarty}, {Chiu}, {Finger},
  {Koh}, {Nelson}, {Prince}, {Rubin}, {Scott}, {Stollberg}, {Vaughan},
  {Wilson}, \& {Wilson}}]{1997ApJS..113..367B}
{Bildsten}, L., {Chakrabarty}, D., {Chiu}, J., {et~al.} 1997, \apjs, 113, 367

\bibitem[{{Blondin} {et~al.}(1990){Blondin}, {Kallman}, {Fryxell}, \&
  {Taam}}]{1990ApJ...356..591B}
{Blondin}, J.~M., {Kallman}, T.~R., {Fryxell}, B.~A., \& {Taam}, R.~E. 1990,
  \apj, 356, 591

\bibitem[{{Bondi}(1952)}]{1952MNRAS.112..195B}
{Bondi}, H. 1952, \mnras, 112, 195

\bibitem[{{Bondi} \& {Hoyle}(1944)}]{1944MNRAS.104..273B}
{Bondi}, H., \& {Hoyle}, F. 1944, \mnras, 104, 273

\bibitem[{{Bozzo} {et~al.}(2016){Bozzo}, {Oskinova}, {Feldmeier}, \&
  {Falanga}}]{2016A&A...589A.102B}
{Bozzo}, E., {Oskinova}, L., {Feldmeier}, A., \& {Falanga}, M. 2016, \aap, 589,
  A102

\bibitem[{{Brookshaw} \& {Tavani}(1993)}]{1993ApJ...410..719B}
{Brookshaw}, L., \& {Tavani}, M. 1993, \apj, 410, 719

\bibitem[{{Calder{\'o}n} {et~al.}(2020){Calder{\'o}n}, {Cuadra}, {Schartmann},
  {Burkert}, {Prieto}, \& {Russell}}]{2020MNRAS.493..447C}
{Calder{\'o}n}, D., {Cuadra}, J., {Schartmann}, M., {et~al.} 2020, \mnras, 493,
  447

\bibitem[{{Castor} {et~al.}(1975){Castor}, {Abbott}, \&
  {Klein}}]{1975ApJ...195..157C}
{Castor}, J.~I., {Abbott}, D.~C., \& {Klein}, R.~I. 1975, \apj, 195, 157

\bibitem[{{Chen} {et~al.}(2018){Chen}, {Blackman}, {Nordhaus}, {Frank}, \&
  {Carroll-Nellenback}}]{2018MNRAS.473..747C}
{Chen}, Z., {Blackman}, E.~G., {Nordhaus}, J., {Frank}, A., \&
  {Carroll-Nellenback}, J. 2018, \mnras, 473, 747

\bibitem[{{Choi} {et~al.}(2016){Choi}, {Dotter}, {Conroy}, {Cantiello},
  {Paxton}, \& {Johnson}}]{2016ApJ...823..102C}
{Choi}, J., {Dotter}, A., {Conroy}, C., {et~al.} 2016, \apj, 823, 102

\bibitem[{{Davidson} \& {Ostriker}(1973)}]{1973ApJ...179..585D}
{Davidson}, K., \& {Ostriker}, J.~P. 1973, \apj, 179, 585

\bibitem[{{De} {et~al.}(2020){De}, {MacLeod}, {Everson}, {Antoni}, {Mandel}, \&
  {Ramirez-Ruiz}}]{2020ApJ...897..130D}
{De}, S., {MacLeod}, M., {Everson}, R.~W., {et~al.} 2020, \apj, 897, 130

\bibitem[{{Dotter}(2016)}]{2016ApJS..222....8D}
{Dotter}, A. 2016, \apjs, 222, 8

\bibitem[{{Duch{\^e}ne} \& {Kraus}(2013)}]{2013ARA&A..51..269D}
{Duch{\^e}ne}, G., \& {Kraus}, A. 2013, \araa, 51, 269

\bibitem[{{Edgar}(2004)}]{2004NewAR..48..843E}
{Edgar}, R. 2004, \nar, 48, 843

\bibitem[{{Eggleton}(1983)}]{1983ApJ...268..368E}
{Eggleton}, P.~P. 1983, \apj, 268, 368

\bibitem[{{El Mellah} {et~al.}(2020{\natexlab{a}}){El Mellah}, {Bolte},
  {Decin}, {Homan}, \& {Keppens}}]{2020A&A...637A..91E}
{El Mellah}, I., {Bolte}, J., {Decin}, L., {Homan}, W., \& {Keppens}, R.
  2020{\natexlab{a}}, \aap, 637, A91

\bibitem[{{El Mellah} {et~al.}(2020{\natexlab{b}}){El Mellah}, {Grinberg},
  {Sundqvist}, {Driessen}, \& {Leutenegger}}]{2020A&A...643A...9E}
{El Mellah}, I., {Grinberg}, V., {Sundqvist}, J.~O., {Driessen}, F.~A., \&
  {Leutenegger}, M.~A. 2020{\natexlab{b}}, \aap, 643, A9

\bibitem[{{El Mellah} {et~al.}(2019){El Mellah}, {Sander}, {Sundqvist}, \&
  {Keppens}}]{2019A&A...622A.189E}
{El Mellah}, I., {Sander}, A.~A.~C., {Sundqvist}, J.~O., \& {Keppens}, R. 2019,
  \aap, 622, A189

\bibitem[{{El Mellah} {et~al.}(2018){El Mellah}, {Sundqvist}, \&
  {Keppens}}]{2018MNRAS.475.3240E}
{El Mellah}, I., {Sundqvist}, J.~O., \& {Keppens}, R. 2018, \mnras, 475, 3240

\bibitem[{{Falanga} {et~al.}(2015){Falanga}, {Bozzo}, {Lutovinov},
  {Bonnet-Bidaud}, {Fetisova}, \& {Puls}}]{2015A&A...577A.130F}
{Falanga}, M., {Bozzo}, E., {Lutovinov}, A., {et~al.} 2015, \aap, 577, A130

\bibitem[{{Friend} \& {Abbott}(1986)}]{1986ApJ...311..701F}
{Friend}, D.~B., \& {Abbott}, D.~C. 1986, \apj, 311, 701

\bibitem[{{F{\"u}rst} {et~al.}(2010){F{\"u}rst}, {Kreykenbohm}, {Pottschmidt},
  {Wilms}, {Hanke}, {Rothschild}, {Kretschmar}, {Schulz}, {Huenemoerder},
  {Klochkov}, \& {Staubert}}]{2010A&A...519A..37F}
{F{\"u}rst}, F., {Kreykenbohm}, I., {Pottschmidt}, K., {et~al.} 2010, \aap,
  519, A37

\bibitem[{{Gies} {et~al.}(2003){Gies}, {Bolton}, {Thomson}, {Huang}, {McSwain},
  {Riddle}, {Wang}, {Wiita}, {Wingert}, {Cs{\'a}k}, \&
  {Kiss}}]{2003ApJ...583..424G}
{Gies}, D.~R., {Bolton}, C.~T., {Thomson}, J.~R., {et~al.} 2003, \apj, 583, 424

\bibitem[{{Hadrava} \& {{\v{C}}echura}(2012)}]{2012A&A...542A..42H}
{Hadrava}, P., \& {{\v{C}}echura}, J. 2012, \aap, 542, A42

\bibitem[{{Hernquist} \& {Katz}(1989)}]{1989ApJS...70..419H}
{Hernquist}, L., \& {Katz}, N. 1989, \apjs, 70, 419

\bibitem[{{Hoyle} \& {Lyttleton}(1939)}]{1939PCPS...35..405H}
{Hoyle}, F., \& {Lyttleton}, R.~A. 1939, Proceedings of the Cambridge
  Philosophical Society, 35, 405

\bibitem[{{Huarte-Espinosa} {et~al.}(2013){Huarte-Espinosa},
  {Carroll-Nellenback}, {Nordhaus}, {Frank}, \&
  {Blackman}}]{2013MNRAS.433..295H}
{Huarte-Espinosa}, M., {Carroll-Nellenback}, J., {Nordhaus}, J., {Frank}, A.,
  \& {Blackman}, E.~G. 2013, \mnras, 433, 295

\bibitem[{{Hurley} {et~al.}(2002){Hurley}, {Tout}, \&
  {Pols}}]{2002MNRAS.329..897H}
{Hurley}, J.~R., {Tout}, C.~A., \& {Pols}, O.~R. 2002, \mnras, 329, 897

\bibitem[{{Jahanara} {et~al.}(2005){Jahanara}, {Mitsumoto}, {Oka}, {Matsuda},
  {Hachisu}, \& {Boffin}}]{2005A&A...441..589J}
{Jahanara}, B., {Mitsumoto}, M., {Oka}, K., {et~al.} 2005, \aap, 441, 589

\bibitem[{{Krti{\v{c}}ka} {et~al.}(2018){Krti{\v{c}}ka}, {Kub{\'a}t}, \&
  {Krti{\v{c}}kov{\'a}}}]{2018A&A...620A.150K}
{Krti{\v{c}}ka}, J., {Kub{\'a}t}, J., \& {Krti{\v{c}}kov{\'a}}, I. 2018, \aap,
  620, A150

\bibitem[{{Lamers} \& {Cassinelli}(1999)}]{1999isw..book.....L}
{Lamers}, H. J.~G.~L.~M., \& {Cassinelli}, J.~P. 1999, {Introduction to Stellar
  Winds}

\bibitem[{{Levine} {et~al.}(1993){Levine}, {Rappaport}, {Deeter}, {Boynton}, \&
  {Nagase}}]{1993ApJ...410..328L}
{Levine}, A., {Rappaport}, S., {Deeter}, J.~E., {Boynton}, P.~E., \& {Nagase},
  F. 1993, \apj, 410, 328

\bibitem[{{Levine} {et~al.}(2000){Levine}, {Rappaport}, \&
  {Zojcheski}}]{2000ApJ...541..194L}
{Levine}, A.~M., {Rappaport}, S.~A., \& {Zojcheski}, G. 2000, \apj, 541, 194

\bibitem[{{Lin}(1977)}]{1977MNRAS.179..265L}
{Lin}, D.~N.~C. 1977, \mnras, 179, 265

\bibitem[{{Lucy} \& {Solomon}(1970)}]{1970ApJ...159..879L}
{Lucy}, L.~B., \& {Solomon}, P.~M. 1970, \apj, 159, 879

\bibitem[{{MacLeod} {et~al.}(2017){MacLeod}, {Antoni}, {Murguia-Berthier},
  {Macias}, \& {Ramirez-Ruiz}}]{2017ApJ...838...56M}
{MacLeod}, M., {Antoni}, A., {Murguia-Berthier}, A., {Macias}, P., \&
  {Ramirez-Ruiz}, E. 2017, \apj, 838, 56

\bibitem[{{MacLeod} {et~al.}(2018){MacLeod}, {Ostriker}, \&
  {Stone}}]{2018ApJ...863....5M}
{MacLeod}, M., {Ostriker}, E.~C., \& {Stone}, J.~M. 2018, \apj, 863, 5

\bibitem[{{Miller-Jones}(2020)}]{2020AAS...23535504M}
{Miller-Jones}, J.~C.~A. 2020, in American Astronomical Society Meeting
  Abstracts, Vol. 235, American Astronomical Society Meeting Abstracts \#235,
  355.04

\bibitem[{{Miller-Jones} {et~al.}(2021){Miller-Jones}, {Bahramian}, {Orosz},
  {Mandel}, {Gou}, {Maccarone}, {Neijssel}, {Zhao}, {Zi{\'o}{\l}kowski},
  {Reid}, {Uttley}, {Zheng}, {Byun}, {Dodson}, {Grinberg}, {Jung}, {Kim},
  {Marcote}, {Markoff}, {Rioja}, {Rushton}, {Russell}, {Sivakoff}, {Tetarenko},
  {Tudose}, \& {Wilms}}]{2021Sci...371.1046M}
{Miller-Jones}, J. C.~A., {Bahramian}, A., {Orosz}, J.~A., {et~al.} 2021,
  Science, 371, 1046

\bibitem[{{Mohamed} \& {Podsiadlowski}(2011)}]{2011ASPC..445..355M}
{Mohamed}, S., \& {Podsiadlowski}, P. 2011, in Astronomical Society of the
  Pacific Conference Series, Vol. 445, Why Galaxies Care about AGB Stars II:
  Shining Examples and Common Inhabitants, ed. F.~{Kerschbaum}, T.~{Lebzelter},
  \& R.~F. {Wing}, 355

\bibitem[{{M{\"u}ller} \& {Vink}(2008)}]{2008A&A...492..493M}
{M{\"u}ller}, P.~E., \& {Vink}, J.~S. 2008, \aap, 492, 493

\bibitem[{{Paxton} {et~al.}(2011){Paxton}, {Bildsten}, {Dotter}, {Herwig},
  {Lesaffre}, \& {Timmes}}]{2011ApJS..192....3P}
{Paxton}, B., {Bildsten}, L., {Dotter}, A., {et~al.} 2011, \apjs, 192, 3

\bibitem[{{Paxton} {et~al.}(2013){Paxton}, {Cantiello}, {Arras}, {Bildsten},
  {Brown}, {Dotter}, {Mankovich}, {Montgomery}, {Stello}, {Timmes}, \&
  {Townsend}}]{2013ApJS..208....4P}
{Paxton}, B., {Cantiello}, M., {Arras}, P., {et~al.} 2013, \apjs, 208, 4

\bibitem[{{Paxton} {et~al.}(2015){Paxton}, {Marchant}, {Schwab}, {Bauer},
  {Bildsten}, {Cantiello}, {Dessart}, {Farmer}, {Hu}, {Langer}, {Townsend},
  {Townsley}, \& {Timmes}}]{2015ApJS..220...15P}
{Paxton}, B., {Marchant}, P., {Schwab}, J., {et~al.} 2015, \apjs, 220, 15

\bibitem[{{Poniatowski} {et~al.}(2021){Poniatowski}, {Sundqvist}, {Kee},
  {Owocki}, {Marchant}, {Decin}, {de Koter}, {Mahy}, \&
  {Sana}}]{2021A&A...647A.151P}
{Poniatowski}, L.~G., {Sundqvist}, J.~O., {Kee}, N.~D., {et~al.} 2021, \aap,
  647, A151

\bibitem[{{Puls} {et~al.}(2008){Puls}, {Vink}, \&
  {Najarro}}]{2008A&ARv..16..209P}
{Puls}, J., {Vink}, J.~S., \& {Najarro}, F. 2008, \aapr, 16, 209

\bibitem[{{Rodr{\'\i}guez-Gonz{\'a}lez}
  {et~al.}(2008){Rodr{\'\i}guez-Gonz{\'a}lez}, {Esquivel}, {Raga}, \&
  {Cant{\'o}}}]{2008ApJ...684.1384R}
{Rodr{\'\i}guez-Gonz{\'a}lez}, A., {Esquivel}, A., {Raga}, A.~C., \&
  {Cant{\'o}}, J. 2008, \apj, 684, 1384

\bibitem[{{Saladino} \& {Pols}(2019)}]{2019A&A...629A.103S}
{Saladino}, M.~I., \& {Pols}, O.~R. 2019, \aap, 629, A103

\bibitem[{{Saladino} {et~al.}(2019){Saladino}, {Pols}, \&
  {Abate}}]{2019A&A...626A..68S}
{Saladino}, M.~I., {Pols}, O.~R., \& {Abate}, C. 2019, \aap, 626, A68

\bibitem[{{Saladino} {et~al.}(2018){Saladino}, {Pols}, {van der Helm},
  {Pelupessy}, \& {Portegies Zwart}}]{2018A&A...618A..50S}
{Saladino}, M.~I., {Pols}, O.~R., {van der Helm}, E., {Pelupessy}, I., \&
  {Portegies Zwart}, S. 2018, \aap, 618, A50

\bibitem[{{Shima} {et~al.}(1985){Shima}, {Matsuda}, {Takeda}, \&
  {Sawada}}]{1985MNRAS.217..367S}
{Shima}, E., {Matsuda}, T., {Takeda}, H., \& {Sawada}, K. 1985, \mnras, 217,
  367

\bibitem[{{Smith}(2014)}]{2014ARA&A..52..487S}
{Smith}, N. 2014, \araa, 52, 487

\bibitem[{{Stone} {et~al.}(2008){Stone}, {Gardiner}, {Teuben}, {Hawley}, \&
  {Simon}}]{2008ApJS..178..137S}
{Stone}, J.~M., {Gardiner}, T.~A., {Teuben}, P., {Hawley}, J.~F., \& {Simon},
  J.~B. 2008, \apjs, 178, 137

\bibitem[{{Stone} {et~al.}(2020){Stone}, {Tomida}, {White}, \&
  {Felker}}]{2020ApJS..249....4S}
{Stone}, J.~M., {Tomida}, K., {White}, C.~J., \& {Felker}, K.~G. 2020, \apjs,
  249, 4

\bibitem[{{Sugimoto} {et~al.}(2017){Sugimoto}, {Kitamoto}, {Mihara}, \&
  {Matsuoka}}]{2017PASJ...69...52S}
{Sugimoto}, J., {Kitamoto}, S., {Mihara}, T., \& {Matsuoka}, M. 2017, \pasj,
  69, 52

\bibitem[{{van Kerkwijk} {et~al.}(1995){van Kerkwijk}, {van Paradijs},
  {Zuiderwijk}, {Hammerschlag-Hensberge}, {Kaper}, \&
  {Sterken}}]{1995A&A...303..483V}
{van Kerkwijk}, M.~H., {van Paradijs}, J., {Zuiderwijk}, E.~J., {et~al.} 1995,
  \aap, 303, 483

\bibitem[{{Vink}(2018)}]{2018A&A...615A.119V}
{Vink}, J.~S. 2018, \aap, 615, A119

\bibitem[{{Vink} {et~al.}(2015){Vink}, {Heger}, {Krumholz}, {Puls}, {Banerjee},
  {Castro}, {Chen}, {Chen{\`e}}, {Crowther}, {Daminelli}, {Gr{\"a}fener},
  {Groh}, {Hamann}, {Heap}, {Herrero}, {Kaper}, {Najarro}, {Oskinova},
  {Roman-Lopes}, {Rosen}, {Sander}, {Shirazi}, {Sugawara}, {Tramper},
  {Vanbeveren}, {Voss}, {Wofford}, \& {Zhang}}]{2015HiA....16...51V}
{Vink}, J.~S., {Heger}, A., {Krumholz}, M.~R., {et~al.} 2015, Highlights of
  Astronomy, 16, 51

\bibitem[{{Woosley} \& {Heger}(2015)}]{2015ASSL..412..199W}
{Woosley}, S.~E., \& {Heger}, A. 2015, {The Deaths of Very Massive Stars}, ed.
  J.~S. {Vink}, Vol. 412, 199

\bibitem[{{Xu} \& {Stone}(2019)}]{2019MNRAS.488.5162X}
{Xu}, W., \& {Stone}, J.~M. 2019, \mnras, 488, 5162

\end{thebibliography}
\end{scriptsize}

\end{document}